\def\ref{\par\noindent\hang}

\def\spose#1{\hbox to 0pt{#1\hss}}
\def\approxlt{\mathrel{\spose{\lower 3pt\hbox{$\sim$}}
        \raise 2.0pt\hbox{$<$}}}
\def\approxgt{\mathrel{\spose{\lower 3pt\hbox{$\sim$}}
        \raise 2.0pt\hbox{$>$}}}

\def\multleft#1{\hbox to size{\vbox {\halign {\lft{##}\cr #1}}\hfill}\par}
\def\multright#1{\hbox to size{\vbox {\halign {\rt{##}\cr #1}}\hfill}\par}

\def\degmark{^\circ}
\def\today{\ifcase\month\or January\or February\or March\or April\or May\or
      June\or July\or August\or September\or October\or November\or December\fi
      \space\number\day, \number\year}
\def\$<${\thinspace}
\def\s{\hbox{\phantom{5}}}      

\def\boxit#1{\vbox{\hrule\hbox{\vrule\kern3pt\vbox{\kern3pt
          #1 \kern3pt}\kern3pt\vrule}\hrule}}

\def\cm{{\rm\thinspace cm}}

\def\erg{{\rm\thinspace erg}}

\def\g{{\rm\thinspace g}}

\def\K{{\rm\thinspace K}}
\def\keV{{\rm\thinspace keV}}
\def\km{{\rm\thinspace km}}

\def\Msun{\hbox{$\rm\thinspace M_{\odot}$}}
\def\pc{{\rm\thinspace pc}}

\def\s{{\rm\thinspace s}}
\def\yr{{\rm\thinspace yr}}

\def\Hz{{\rm\thinspace Hz}}

\def\pcmcu{\hbox{$\cm^{-3}\,$}}

\def\ergpcmsqps{\hbox{$\erg\cm^{-2}\s^{-1}\,$}}

\def\ergps{\hbox{$\erg\s^{-1}\,$}}

\def\kmps{\hbox{$\km\s^{-1}\,$}}

\def\pcmsq{\hbox{$\cm^{-2}\,$}}

\def\ps{\hbox{$\s^{-1}\,$}}

\documentstyle[psfig]{mn}

\title[A multi-wavelength study of MCG$-$6-30-15]
{A multi-wavelength study of the Seyfert 1 galaxy MCG$-$6-30-15}

\author[C.~S.~Reynolds et al.]
{C.~S.~Reynolds$^{1,2}$, M.~J.~Ward$^3$, A.~C.~Fabian$^2$ and A.~Celotti$^4$\\
{\small $^1$JILA, University of Colorado, Boulder, Colorado, CO~80309-0440,
USA}\\
{\small $^2$Institute of Astronomy, Madingley Road, Cambridge, CB3 0HA}\\
{\small $^3$X-ray Astronomy Group, Department of Physics and Astronomy,
University of Leicester, Leicester, LE1 7RH}\\
{\small $^4$S.I.S.S.A., via Beirut 2-4, 34014 Trieste, Italy}\\
}

\begin{document}

\maketitle

\begin{abstract}
We present a multiwaveband spectroscopic study of the nearby Seyfert 1
galaxy MCG$-$6-30-15.  New optical spectra from the
Anglo-Australian Telescope are presented which clearly show the effects of
dust extinction/reddening on both the emission line spectrum and the
non-stellar AGN continuum.  The reddening is constrained to be in the range
$E(B-V)=0.61-1.09$.  Spectroscopy in the X-ray band, with both {\it ROSAT}
and {\it ASCA}, reveal absorption by the warm absorber but little or no
neutral absorption expected to accompany the dust responsible for the
optical reddening.  The dusty warm absorber solution to this discrepancy is
discussed and photoionization models of such warm absorbers are
constructed.  The optical spectrum also displays the relatively strong
`coronal' lines of [FeX]$\lambda 6375$, [FeXI]$\lambda 7892$ and
[FeXIV]$\lambda 5303$.  We show that these lines may plausibly originate
from the outer regions of the warm absorber, although better calculations
of the collision strengths for these transitions are required in order to
conclusively address this issue.  We also present new ultraviolet data
from the {\it International Ultraviolet Explorer} and suggest that much of the
observed UV flux is scattered into our line of sight (with a scattering
fraction of 1--5 per cent).  We conclude with a discussion of the global
energetics of this system.   
\end{abstract}


\section{Introduction}

Accretion of material onto a supermassive black hole has long been believed
to be the fundamental power source of active galactic nuclei (AGN; e.g. see
Rees 1984 for a review).  However, the physical processes by which the
gravitational potential energy of the accretion flow energizes the observed
spectrum are still far from certain.  Recent spectroscopic studies in the
X-ray waveband have shown that, in the innermost regions (i.e. $r\approxlt
10\,{\rm R}_{\rm Sch}$; where ${\rm R}_{\rm Sch}$ is the Schwarzschild
radius of the central black hole) of at least some AGN, there is a
geometrically-thin, radiatively-efficient accretion disk (Tanaka et
al. 1995; Fabian et al. 1995).  A significant fraction of the accretion
energy appears to be liberated in a hot, optically-thin, disk-corona which
is a prolific radiator of X-rays and $\gamma$-rays (probably via the
Comptonization of optical/UV photons from the cold accretion disk: Haardt
\& Maraschi 1991; Field \& Rogers 1993; Zdziarski et al. 1994).  Although
most of the primary radiation is produced in the inner regions of the
accretion flow, a significant fraction of this radiation is reprocessed at
greater distances from the black hole into UV, optical and IR wavelengths.
Studying these reprocessing mechanisms allows the structures surrounding
the accreting black hole to be probed and are necessary if we are to
disentangle the primary emission from the reprocessed emission.

In recent years, much has been learnt about the various reprocessing
mechanisms.  It has been realized that approximately half of the X-ray
photons emitted from the corona will strike the cold accretion disk,
thereby producing `reflection' features in the X-ray spectrum (Guilbert \&
Rees 1988; Lightman \& White 1988; Pounds et al. 1990).  The emission that
emerges from these central regions can also be intercepted by more distant
structures: these include the broad-line region (BLR), the putative
molecular torus of the unified Seyfert schemes and the warm absorber.  The
scattering of primary radiation into our line-of-sight is also known to be
important in at least some AGN.  Multi-waveband studies of nearby,
bright AGN are invaluable in studying such complex systems.

Many such multi-wavelength studies of AGN have been performed.  For
example, Alloin et al. (1995) utilized many ground-based and space
observatories to obtain a snapshot (i.e. all data taken almost
simultaneously) of the Seyfert 1 galaxy NGC~3783.  After careful
consideration of possible contaminating sources, these authors fit thermal
accretion disk models to the classical big blue bump displayed by this
object and hence constrain the black hole mass and accretion rate.  They
also find an infrared bump which they interpret as thermal emission from
hot and warm dust.  Many other studies focus on multi-wavelength monitoring
in an attempt to map the central engine.  For example,
extensive monitoring campaigns have been performed on NGC~4151 (Edelson et
al. 1996) and NGC~5548 (Korista et al. 1995).

In this paper we perform a multiwaveband study of the nearby Seyfert 1
galaxy MCG$-$6-30-15 ($z=0.008$; Pineda et al. 1978; Pineda et al. 1980).
Snapshot optical imaging with the {\it Hubble Space Telescope} (HST)
clearly shows this object to be a S0-type galaxy with a bright, nuclear
point source.  This X-ray bright AGN has recently been the subject of
intense spectroscopic study in the X-ray band and so is a good candidate
for a multiwavelength study.  We present new optical data from the
Anglo-Australian Telescope (AAT) and new ultraviolet data from the {\it
International Ultraviolet Explorer} satellite (IUE).  Together with
archival infrared and X-ray data, this represents the most detailed
multiwaveband study of this nearby AGN to date.  In particular, we review
and reinforce evidence that there is a significant column of dusty ionized
material along our line of sight to the central continuum source and BLR,
the so-called dusty warm absorber.  This material is shown to have a major
effect on the observed spectrum of this source.

In Section 2, the data are presented and the basic characteristics at each
waveband are discussed.  Section 3 examines various estimates for the
amount of extinction and absorption towards the central source.  A
discrepancy between the optical/UV extinction and the X-ray absorption
leads us to discuss the possibility of dusty warm absorbers in Section 4.
Section 5 probes the connection between the warm absorber and coronal line
emission.  We discuss our results and their implications for the energetics
of the system in Section 6.  Section 7 presents our conclusions.

\section{The Data and basic spectral properties}

\subsection{Optical}

\begin{figure*}
\hbox{
\hspace{2cm}
\psfig{figure=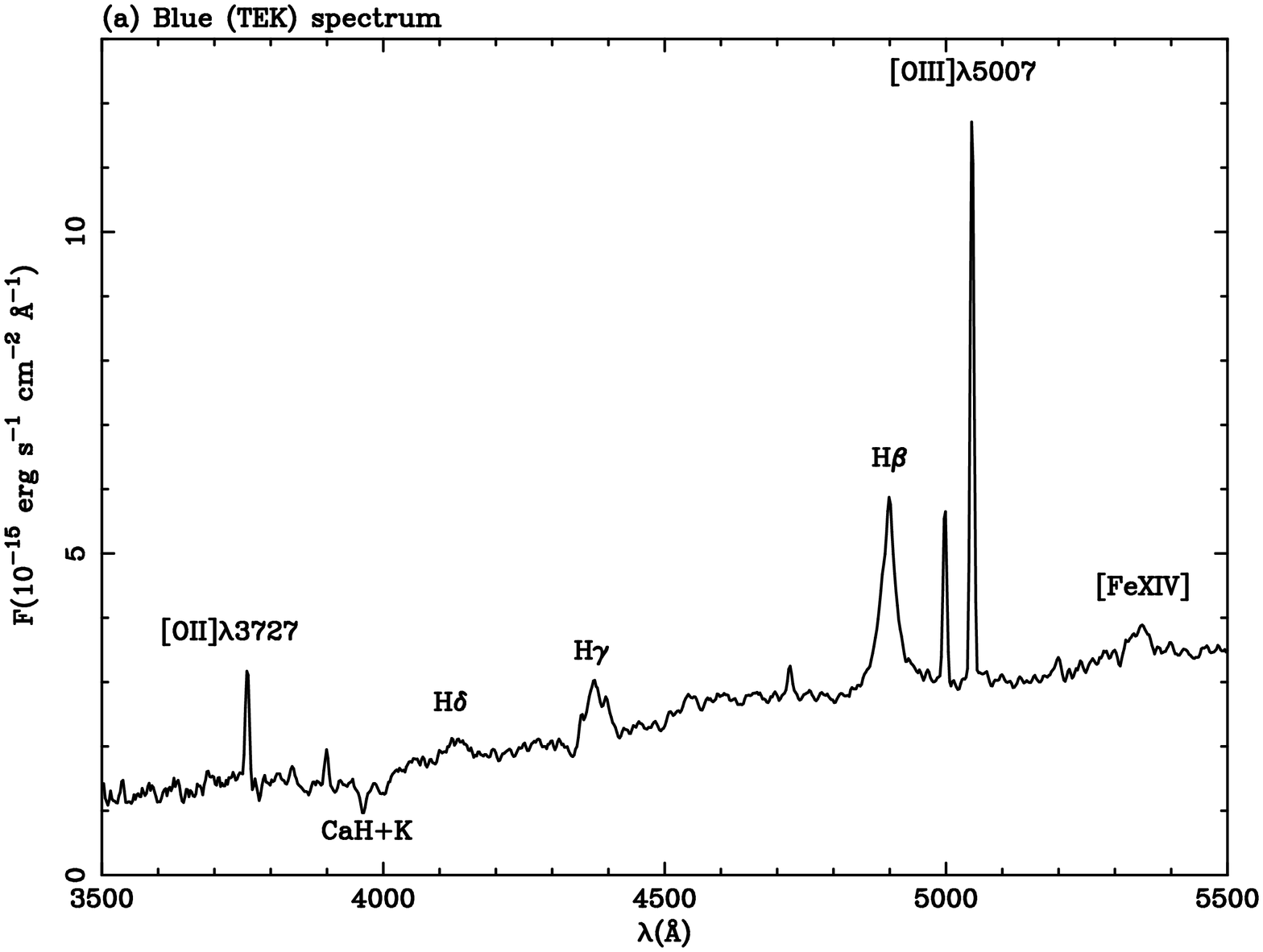,width=0.8\textwidth,angle=270}
}
\hbox{
\hspace{2cm}
\psfig{figure=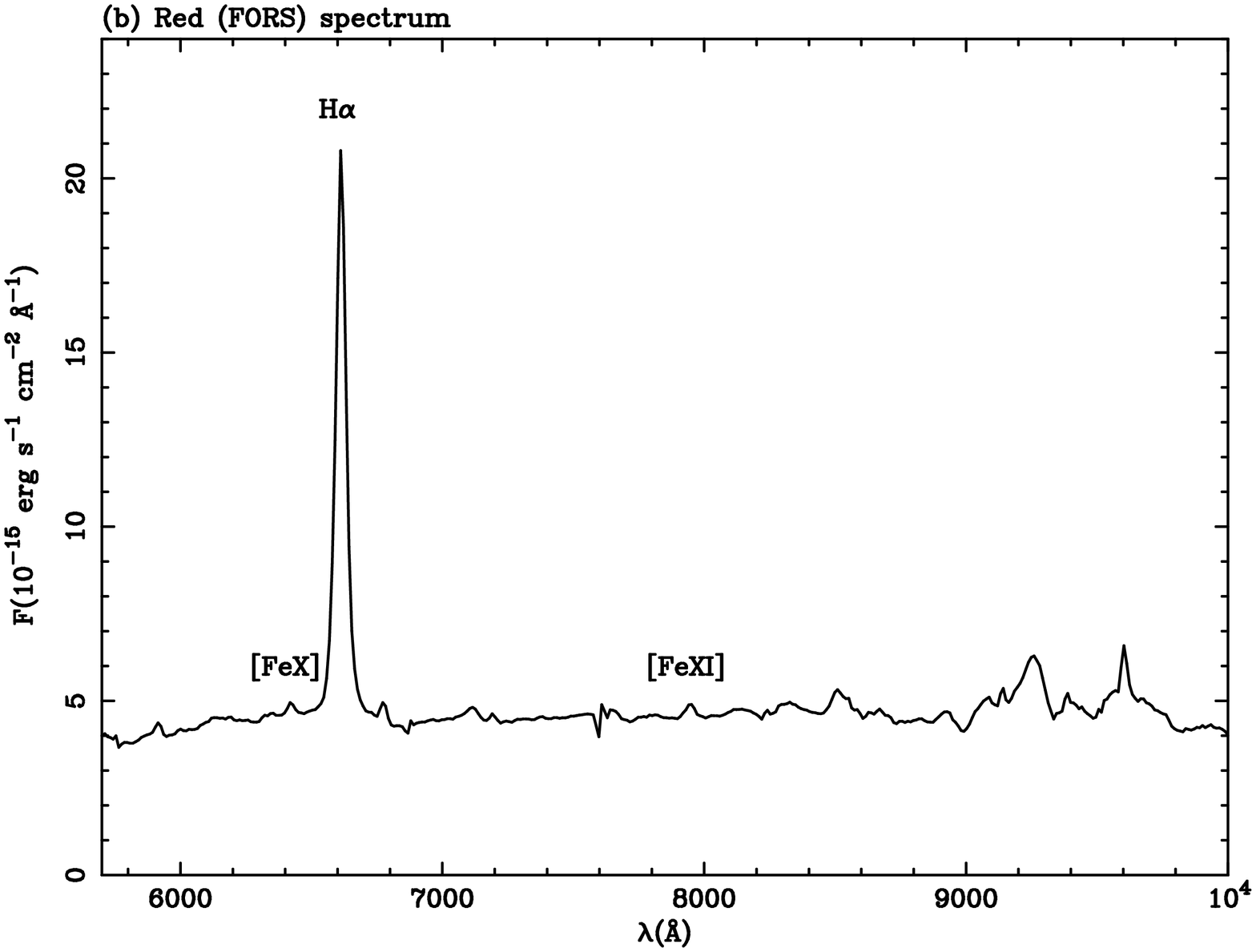,width=0.8\textwidth,angle=270}
}
\caption{The blue (a) and red (b) optical spectra of the nuclear region of
MCG$-$6-30-15.  Prominent emission lines and the Ca\,{\sc ii} absorption
feature have been marked.}
\end{figure*}

Optical spectra of MCG$-$6-30-15 were obtained with the AAT on the
night of 1995 April 10.  The RGO spectrograph with the 300B grating
(grating angle $23.27\degmark$) was used to obtain blue (3500--5500\AA\,)
spectra and the Faint Object Red Spectrograph (FORS) provided red
(5700--10000\AA\,) spectra.  Red and blue spectra could be taken
simultaneously\footnote{Due to the shutter configuration and the read-out
constraints, the red spectral integrations had to be terminated and
read-out 60\,s before the blue.} via the use of a dichroic beam splitter.
The spectral (FWHM) resolutions are $\sim 9$\AA\, in the blue and $\sim
30$\AA\, in the red.  Standard data reduction was performed using the {\sc
figaro} software package provided by {\sc starlink}.  Wavelength
calibration was performed using a Cu-Ar arc.  We estimate that the
wavelength calibration is accurate to within 1\AA\ in the blue spectrum and
3\AA\ in the red spectrum.  Absolute flux calibration utilized the standard
star L745-46A.  Finally, the atmospheric water bands within the red
spectrum were corrected for using the extremely metal-deficient red giant
star HD126587 as a continuum reference source.

In order to be sensitive to intra-night variations in flux or spectrum, we
took 18 separate red/blue spectra of MCG$-$6-30-15.  A normal Galactic
F-star is situated 6 arcsec to the south of the nucleus of MCG$-$6-30-15
(Pineda et al. 1980): the slit position was arranged so as to cover both
the nucleus and the star.  Assuming the star to have a constant flux during
the night, this provides a good control against which we can search for
flux variations in the nucleus.  The exposure time of each spectrum was
600\,s in the blue and 540\,s in the red.

After reduction it was found that only 8 red/blue spectra were unaffected
by the variable weather conditions during the night.  The spectrum of the
nuclear region of MCG$-$6-30-15 was extracted from each long-slit spectrum.
No significant difference in either overall flux or individual line fluxes
could be found between these spectra.  Furthermore, no significant
variability could be detected when the overall nuclear flux was compared
with that of the nearby F-star.  These conclusions remain robust even if we
include periods of data that were affected by poor weather.  We can set
upper limits of $\sim 5$ per cent on intra-night variation of the optical
flux of the nucleus.  Given the lack of any detectable variability, the 8
`good' spectra were combined to form a single co-added spectrum with
effective exposure times 4800\,s in the blue and 4360\,s in the red.
Figures 1a and 1b show the blue and red coadded spectra, respectively.

These spectra possess extremely high signal-to-noise and, therefore,
uncertainties are dominated by calibration effects and atmospheric
variations.  Comparing the 8 separate good spectra (i.e. those obtained
prior to co-adding), we estimate that there is a $\sim 10$ per cent
uncertainty in the overall normalization of each spectrum.  These are most
likely due to `grey' variations in atmospheric conditions.  Treating these
uncertainties as independent, the coadded spectra should each have an
uncertainty of approximately 3 per cent in overall normalization.  The
uncertainties in the red and blue spectra should be statistically
independent.

Several features are immediately apparent from these spectra.  The emission
line spectrum is dominated by broad Balmer lines (H$\alpha$, H$\beta$,
H$\gamma$ and H$\delta$) and narrow forbidden oxygen lines ([OII]$\lambda
3727$ and [OIII]$\lambda 4959/5007$).  The emission line spectrum, which is
clearly related to the Seyfert activity, is examined in more detail below
(Section 2.1.2).  The spectrum between 9000\AA\ and 10000\AA\ may be severely
affected by incorrect subtraction of atmospheric water features, and so the
line-like features in this region of the spectrum should be treated with
caution.  The presence of the Ca\,{\sc ii} doublet in absorption (near
4000\,\AA\,) shows there to be a non-negligible fraction of starlight from
the host galaxy contributing to this spectrum.  We now examine this stellar
component.

\subsubsection{The galactic spectrum}

\begin{figure*}
\hbox{
\hspace{2cm}
\psfig{figure=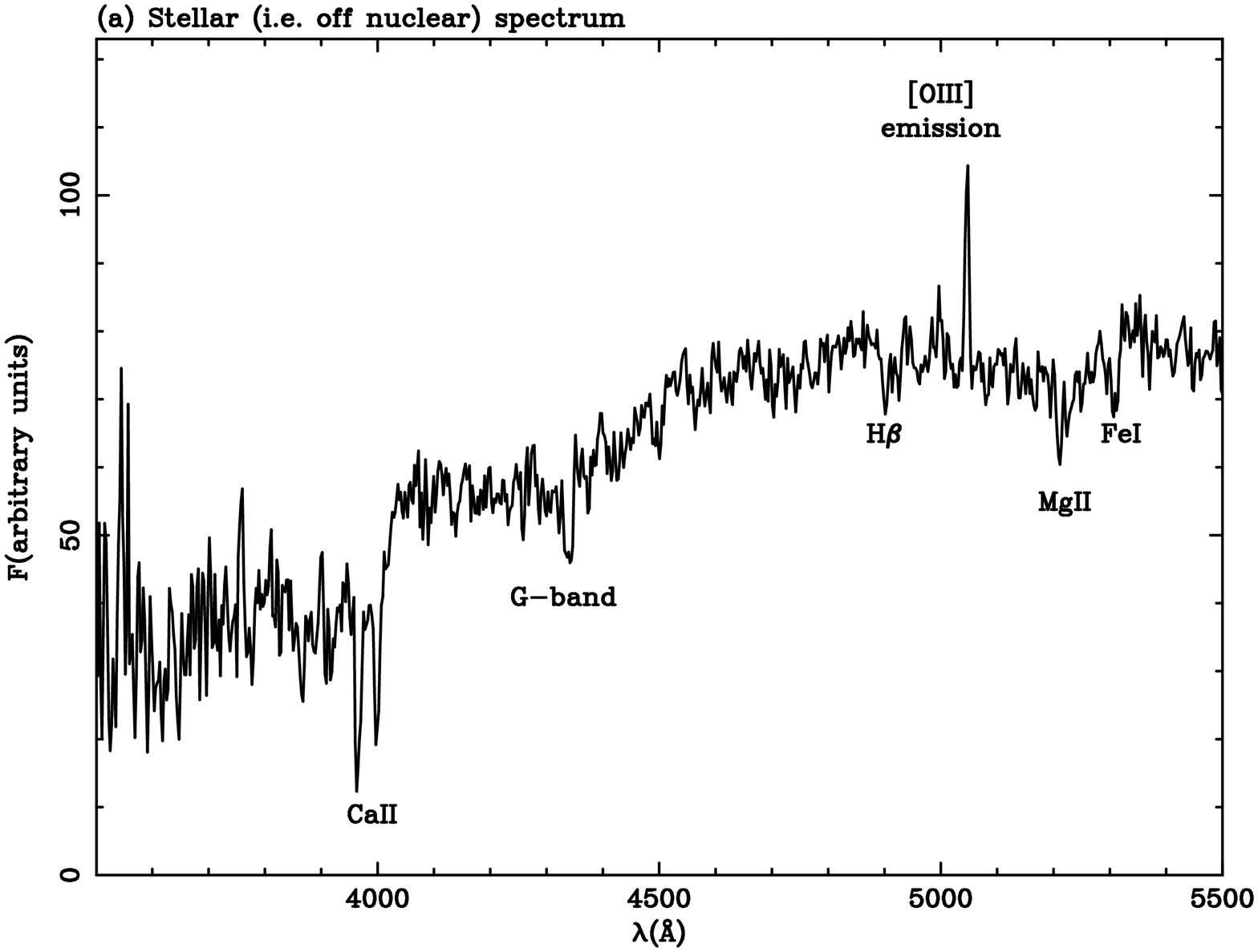,width=0.8\textwidth,angle=270}
}
\hbox{
\hspace{2cm}
\psfig{figure=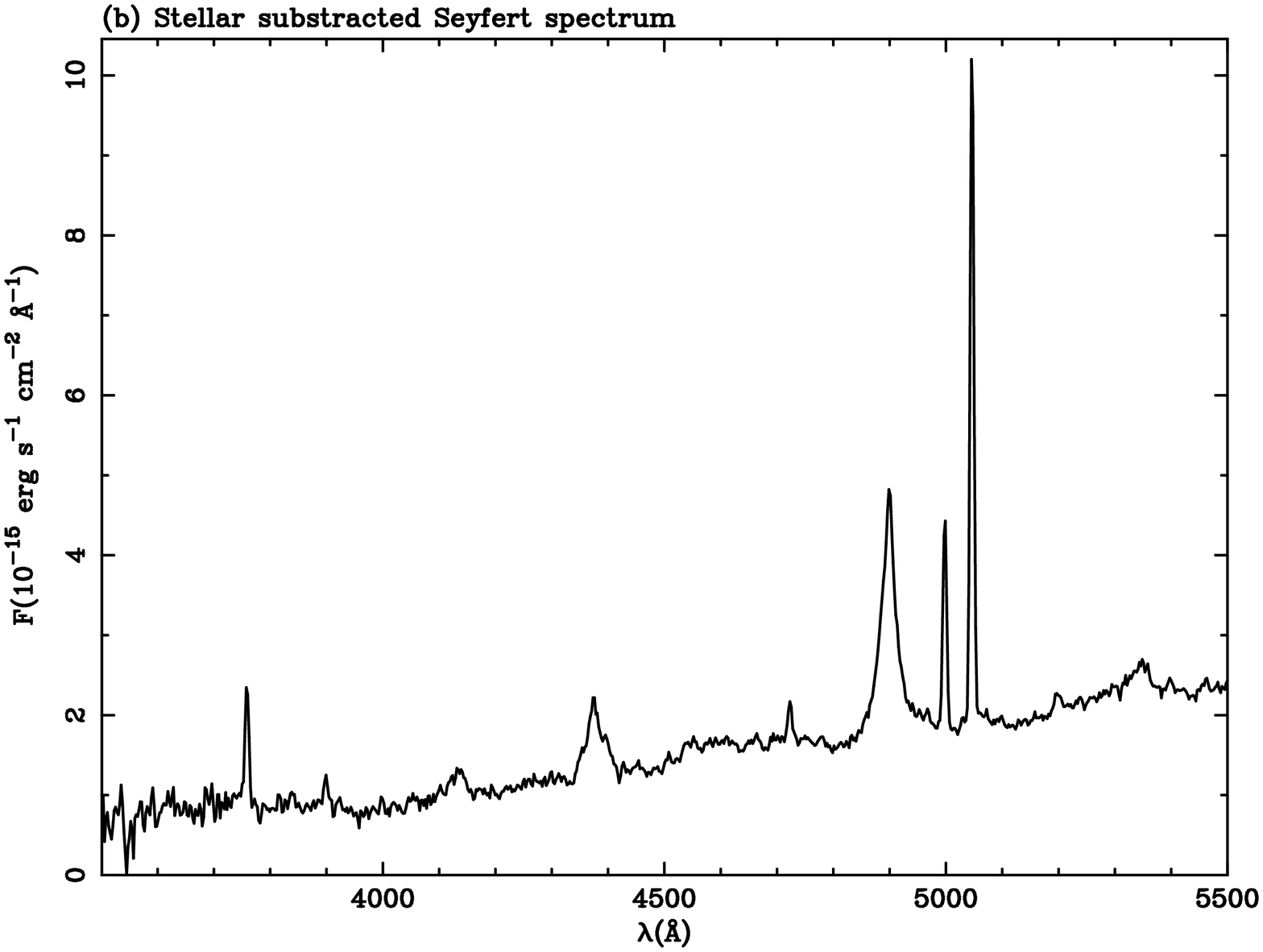,width=0.8\textwidth,angle=270}
}
\caption{a) Spectrum of the off-nuclear, stellar component.  The absence of
broad Balmer lines, together with the strong redshifted Ca\,{\sc ii}
doublet feature verifies that the host galaxy has been successfully
isolated from the nuclear flux.  b) Pure Seyfert spectrum obtained by
subtracting the stellar spectrum from the total nuclear spectrum until the
Ca\,{\sc ii} feature vanishes.  Note that the stellar absorption feature of
Mg~B around 5200\AA\ (rest frame) is also successfully removed.}
\end{figure*}

For the blue data we extracted the spectrum of the host galaxy (which
appears on the long slit frames as a clearly defined `fuzz' surrounding the
bright nuclear emission.)  The spectra extracted from the individual
long-slit frames were again co-added to form a single spectrum.  Figure 2a
shows this co-added spectrum.  The absence of any broad Balmer lines from
this spectrum is verification that we have successfully isolated the
stellar component.  Note that there {\it is} narrow [OII] and [OIII] line
emission present in this spectrum.  We attribute this to the narrow line
region (NLR) possessing significant spatial extent (and therefore
contaminating the spectrum of the host galaxy.)

The stellar spectrum reveals several absorption features.   The weak
H$\beta$ absorption indicates the presence of a young stellar population.
An older stellar population give rise to the Mg~B, Fe\,{\sc i}, G-band and
Ca\,{\sc ii} absorption features.

\subsubsection{The nuclear spectrum}

Assuming the stellar spectrum to be spatially uniform, we can subtract the
spectrum of the host galaxy from the total nuclear spectrum in order to
isolate the spectrum of the active nucleus.  In practice, we progressively
subtract more of the stellar spectrum from the total spectrum until the
Ca\,{\sc ii} doublet feature at $\sim 4000$\AA\, vanishes.  The resulting
difference spectrum is taken to be the spectrum of the active nucleus.
This is shown in Figure 2b.  A similar procedure was not performed for the
red (FORS) data because the poorer spatial resolution of our FORS data
makes isolation of the host galaxy spectrum more difficult.  This would
make subtraction of the stellar component rather subjective and
consequently diminish its value.

\begin{table*}
\begin{center}
\begin{tabular}{lcccccc}\hline
line & rest & observed & relative velocity & FWHM & line flux & equivalent \\
identification & wavelength (\AA) & wavelength (\AA) & of line ($\kmps$) & ($\kmps$)
& ($10^{-15}\ergpcmsqps$) & width (\AA) \\\hline
CIV & 1549 & $1545\pm 6$ & $-3000\pm 1200$ & $6200\pm 2100$ & $250\pm 90$ &
$140\pm 70$ \\
$[{\rm OII}]$  & 3727 & $3758\pm 1$ & $160\pm 80$ & $<790$ & $16\pm 2$ &
$14\pm 1$ \\
$[{\rm NeIII}]$   & 3869 & $3899\pm 1$ & $-10\pm 80$ & $<1030$ & $4.3\pm
1.0$ & $4.0\pm 1.5$ \\
H$\delta$ & 4103 & $4133\pm 1$ & $-140\pm 70$ & $3000\pm 370$ & $18\pm 4$ &
$14\pm 4$ \\
H$\gamma$ & 4342 & $4374\pm 1$ & $-130\pm 70$ & $2470\pm 180$ & $37\pm 3$ &
$30\pm 3$ \\
$[{\rm OIII}]$ & 4363 & $4400\pm 2$ & $210\pm 140$ & $<1470$ & $5.9\pm 1.2$
& $4\pm 1$ \\
HeII      & 4686 & $4722\pm 1$ & $-30\pm 60$ & $<710$ & $5.4\pm 0.6$ &
$2.7\pm 0.3$ \\
H$\beta$  & 4862 & see text & see text & see text & $119\pm 5$ &
$66\pm 3$ \\
$[{\rm OIII}]$ & 4959 & $4997\pm 1$ & $-40\pm 60$ & $<530$ & $24.0\pm 0.8$
& $11.3\pm 0.3$ \\
$[{\rm OIII}]$ & 5007 & $5046\pm 1$ & $\pm 60$ & $<530$ & $83.0\pm 0.7$ &
$37.2\pm 0.3$ \\
$[{\rm FeVIII}]$ & 5159 & $5199\pm 1$ & $-10\pm 60$ & $860\pm 160$ &
$3.4\pm 0.7$ & $1.4\pm 0.3$ \\
$[{\rm FeXIV}]$ & 5303 & $5343\pm 1$ & $-90\pm 60$ & $2320\pm 300$ & $14\pm
1$ & $5.2\pm 0.4$ \\
$[{\rm FeX}]$ & 6375 & $6422\pm 3$ & $-130\pm 140$ & $1950\pm 500$ & $15\pm
3$ & $3.0\pm 0.4^*$ \\
H$\alpha$ & 6563 & $6613\pm 2$ & $-50\pm 90$ & $2590\pm 90$ & $952\pm 23$ &
$211\pm 20^*$ \\
$[{\rm SII}]$ & 6712 & $6775\pm 3$ & $480\pm 130$ & $<1580$ & $16\pm
3$ & $3.0\pm 0.7^*$\\
$[{\rm FeXI}]$ & 7892 & $7947\pm 3$ & $-250\pm 110$ & $<1850$ &
$15\pm 3$ & $3.0\pm 0.7^*$\\\hline
\end{tabular}
\caption{UV/Optical/NIR line spectrum for MCG$-$6-30-15.  Column 4 shows
the velocity redshift of the line centre with respect to the reference
frame defined by the [OIII]$\lambda 5007$ emission ($z=0.00779$).  Column 6
shows the total flux in the line, corrected for Galactic extinction.  Those
equivalent widths marked with an asterix have been measured from FORS data
which have {\it not} been galaxy-subtracted.  All errors and limits are
quoted at the 1-$\sigma$ level.}
\end{center}
\end{table*}

As is typical for Seyfert 1 nuclei, the spectrum consists of a strong
non-stellar continuum, broad Balmer lines and narrow permitted and
forbidden lines.  Both the red and blue spectra were visually examined for
known prominent lines.  All such identified lines were characterized by
fitting a single Gaussian profile whilst modeling the local continuum as a
power-law.  This procedure was performed on the galaxy-subtracted spectrum,
with the exception of those few lines that were identified in the FORS data
(for the reasons given above).  Since our stellar spectrum is contaminated
with forbidden oxygen line emission (presumably from an extended NLR), we
note that these oxygen lines will be suppressed by $\sim 10$ per cent in
the galaxy-subtracted AGN spectrum.  The single Gaussian parameterization
is a (visually) good fit to all of the emission lines except H$\beta$.
Three Gaussian components are required to properly describe this line:

a) A narrow line component at the systemic velocity of the galaxy (defined as
the velocity of the [OIII] emission line region) with FWHM$\approxlt 700\kmps$
and flux $F_{\rm n}(H\beta)=9.6\times 10^{-15}\ergpcmsqps$.

b) A broad line component blueshifted by $200\pm 60\kmps$ relative to the
systemic velocity with FWHM$\sim 2400\pm 200\kmps$ and flux $F_{\rm
b}(H\beta)=6.3\times 10^{-14}\ergpcmsqps$.

c) A very broad line component redshifted by $500\pm 120\kmps$ relative to the
systemic velocity with FWHM$\sim 7100\pm 700\kmps$ and flux $F_{\rm
vb}(H\beta)=4.6\times 10^{-14}\ergpcmsqps$. 

The resulting line identifications, wavelengths, line widths and total line
fluxes for all of the identified lines are given in Table~1.  The errors
quoted in this table (and those for the H$\beta$ components above) include
both statistical errors and an estimate of any systematic errors resulting
from the wavelength/flux calibration.  The statistical errors are derived
from $\chi^2$ fitting of the Gaussian models to the data, assuming that the
pixel-to-pixel variation is due to random Gaussian noise.

The optical spectrum clearly shows the effect of dust extinction:
the continuum flux declines towards the blue end of the spectrum and the
Balmer decrements are large.  The line-of-sight extinction, and a
comparison of this extinction to the line-of-sight X-ray absorption will be
addressed in Section 3.

\subsection{Ultraviolet}

The ultraviolet data were taken by IUE on 1994 July 23--24 (i.e. three
days prior to the {\it ASCA} observation described below.)  Due to excessive
scattered solar light, we present only the data taken with the short
wavelength primary (SWP) camera.  These short-wavelength data were taken
during one full shift on 1994 July 24 (effective exposure time 8 hours).
Data reduction was performed using the {\sc ark} software package.

\begin{figure}
\hbox{
\hspace{1cm}
\centerline{\psfig{figure=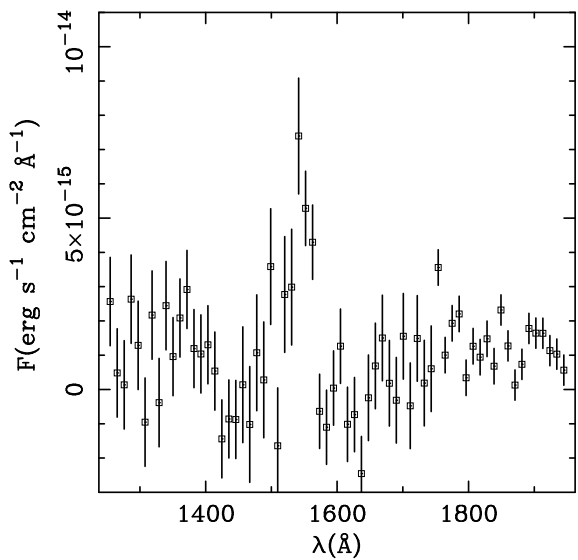,width=0.65\textwidth,angle=270}}
}
\caption{IUE (SWP) data for MCG$-$6-30-15.  Note the weak continuum and the
prominent {\sc C\,iv}$\lambda 1549$ line.}
\end{figure}

The resulting SWP spectrum is shown in Fig.~3.  The source is found to be
weak in the UV compared with other wavebands (apart from radio).  The
hypothesis that this UV spectrum is flat and featureless can be rejected
with a high level of confidence: such a hypothesis gives $\chi^2=153$ for
65 degrees of freedom (dof).  The most significant deviation from the
constant model is clearly the {\sc C\,iv}$\lambda 1549$ resonance emission
line doublet.  Modeling this with a Gaussian profile leads to a highly
significant improvements in the goodness of fit ($\Delta\chi^2=35$ for 3
additional dof).  The inferred properties of this line are reported in
Table~1.  The {\sc C\,iv}$\lambda 1549$ line appears to be significantly
broader than the H$\alpha$ line, and has a FWHM comparable with the
`very-broad' component of the H$\beta$ line.  The {\sc C\,iv} line is also
blueshifted by $\sim 3000\kmps$ with respect to the broad Balmer lines.
This appears to be a generic feature of high-ionization lines from AGN
(e.g. see Espey et al. 1989).  This has been interpreted by some authors as
evidence for a two-component BLR (e.g. Collin-Souffrin et al. 1988).

There may also be an absorption trough to the blueside of the {\sc
C\,iv}$\lambda 1549$ lines.  Modeling this with a Gaussian profile leads
to a further improvement in the goodness of fit by $\Delta\chi^2=12$ for 3
additional dof.  According to the F-test, this is not a significant
improvement at the 90 per cent level.  Thus, we cannot conclusively
determine the reality of this feature and shall not discuss it further.

Formally, the best fit continuum level is $F_{\rm SWP}=(1.1\pm 0.2)\times
10^{-15}\ergpcmsqps\AA^{-1}$.  It is possible that much of this UV
continuum could be stellar in origin.  Thus, this value should be
considered only as an upper limit to the UV continuum flux from the AGN.
High-resolution UV imaging with HST will allow the stellar UV flux to be
almost completely separated from the AGN UV flux, thereby allowing these
two components to be separated.  At the time of writing, such UV imaging
has yet to be performed.

\subsection{Other wavebands}

To make this study genuinely multi-waveband in nature, we have supplemented
the above new spectra will previously published data.   The remainder of this
section will introduce these data.

\subsubsection{Infrared}

Mid/far infrared (MIR/FIR) data were obtained from the {\it Infrared
Astronomy Satellite} (IRAS) Faint Source Catalogue (version 2.0) at four
wavelengths ($12\,\mu{\rm m}$, $25\,\mu{\rm m}$, $60\,\mu{\rm m}$ and
$100\,\mu{\rm m}$.)  Giuricin, Marddirossian \& Mezzetti (1995) find good
agreement between the IRAS $12\,\mu{\rm m}$ flux and the ground-based,
small-beam $10\,\mu{\rm m}$ (N-band) flux.  Thus, it appears there are no
strong confusing IR sources in the IRAS beam at the short IRAS wavelengths.
Unless there is a very cool confusing source which only reveals itself at
the longer IRAS wavelengths, this suggests that IRAS is providing a
reliable measurement of the flux of MCG$-$6-30-15 at all IR wavelengths.
Furthermore, this agreement between the ground-based observations and IRAS
implies that there is little contribution to the NIR flux from the host
galaxy.  This is consistent with it being of Hubble type S0.

J-, H-, K- and L-band (i.e. $1.2-3.5\mu{\rm m}$) fluxes were obtained from
the ground-based work of Ward et al. (1987).

\subsubsection{Soft (0.2--2\,keV) X-rays}

Soft X-ray data from the {\it ROSAT} Position Sensitive Proportional
Counter (PSPC) were retrieved from the public archive situated at the NASA
Goddard Space Flight Center.  We chose the longest of the two available
PSPC observations of this object.  This observation started on
1992-January-29 and collected a total of 8500\,s of good data.  A PSPC
spectrum of MCG$-$6-30-15 was formed by binning all counts extracted from a
circular region of radius 2\,arcmin centered on the point source
corresponding to MCG$-$6-30-15.  The background count rate within this
region is found to be negligible compared with the source.

These data have been previously discussed by Nandra \& Pounds (1992).
These authors found evidence for a notch in the PSPC spectrum at an energy
of $E\sim 0.8\keV$.  They interpreted this feature as a blend of {\sc
O\,vii} and {\sc O\,viii} K-shell absorption edges resulting from
highly-ionized material along the line of sight to the primary (power-law)
X-ray source.  This material has become known as the {\it warm absorber}
(Halpern 1984), and shall play a major role in the discussion presented in
this paper.

\subsubsection{Hard (0.6--10\,keV) X-rays}

The Advanced Satellite for Cosmology and Astrophysics ({\it ASCA})
performed a long (4.5-day) observation of MCG$-$6-30-15 on 1994 July 27.
The `good' exposure times were $\sim 1.5\times 10^5\s$ in each of the four
detectors.  As a result, high signal-to-noise, medium-resolution X-ray
spectra were obtained in the 0.6--10\,keV band.  The fact that the X-ray
flux is highly variable on short timescales, together with {\it ROSAT}
imaging, confirms that the observed X-rays are originating from the central
engine of this Seyfert nucleus (see Fig.~1 of Fabian et al. 1995).

Detailed analyses of these data are presented in Tanaka et al. (1995),
Otani et al. (1996) and Iwasawa et al. (1996).  As discussed by these
authors, the 0.6--10\,keV X-ray spectrum shows clear deviations from the
canonical power-law form.  In particular, the {\sc O\,vii} and {\sc
O\,viii} edges from the warm absorber are prominent features in the
0.7--2\,keV range and a broad emission feature is seen between 5--7\,keV.

{\it ASCA} revealed the warm absorber in MCG$-$6-30-15 to be highly variable
(Fabian et al. 1994; Reynolds et al. 1995; Otani et al. 1996).   Detailed
modeling of this variability (Otani et al. 1996; Reynolds 1996)  has led to a
two-zone model for this absorber.  There seems to be a highly
photoionized inner region (possibly related to the BLR) and a less ionized
outer region (possibly related to the putative molecular torus or NLR).  It
is likely that both of these absorbers are in outflow driven by the
radiation pressure of the central source (Reynolds \& Fabian 1995).  

The emission feature at 5--7\,keV is thought to be due to the fluorescent
K$\alpha$ line emission of cold iron (i.e. Fe\,{\sc i}--Fe\,{\sc xvii})
that results when primary X-rays illuminate cold material in the vicinity
of the central engine (George \& Fabian 1991; Matt, Perola \& Piro 1991).
The rest-frame energy of this emission line is 6.4\,keV.  {\it ASCA}
resolves this line and allows its profile to be determined.  It is found
that the profile is in good agreement with the hypothesis that it
originates from the innermost regions of a thin, radiatively-efficient,
accretion disk about a black hole (Tanaka et al. 1995; Fabian et al. 1995;
Iwasawa et al. 1996).  Relativistic beaming, transverse Doppler shifts and
gravitational redshifts are of major importance in determining the profile
of this emission line.  From the point of view of the current work, the
iron line observation is important since it constrains the geometry of the
energetically important inner accretion disk: in the innermost region, much
of the energy appears to be liberated as X-rays in an optically-thin region
near a radiatively-efficient thin accretion disk viewed at an inclination
of $30\degmark$.  

\subsection{The multi-waveband spectrum}

The above data are compiled together into a multi-waveband spectrum in
Fig.~4.  The spectral data plotted in this figure have not been
stellar-subtracted or corrected for any reddening (i.e. this is the total
galaxy+AGN spectrum as observed at Earth).  The dotted line sketches the
approximate intrinsic (i.e. dereddened/unabsorbed) spectrum of this
source given a reddening of $E(B-V)=0.61$ (see Section 3).  

There are several noteworthy features in this multiwaveband spectrum.
First, the observed spectral energy distribution [i.e. $\nu F(\nu)$] peaks
in the mid-infrared ($\nu\sim 2-3\times 10^{13}\Hz$).  As discussed below,
this is probably due to thermal emission from warm/hot dust grains.
Secondly, whereas many classical Seyfert 1 galaxies display a strong
optical/UV continuum (the so-called Big Blue Bump), MCG$-$6-30-15 is
heavily reddened in the optical band and almost extinguished in the UV.
There is then a strong recovery at X-ray frequencies.  The effect of the
warm absorber and iron K$\alpha$ emission line are clearly seen in the
X-ray spectrum.

\begin{figure*}
\centerline{\psfig{figure=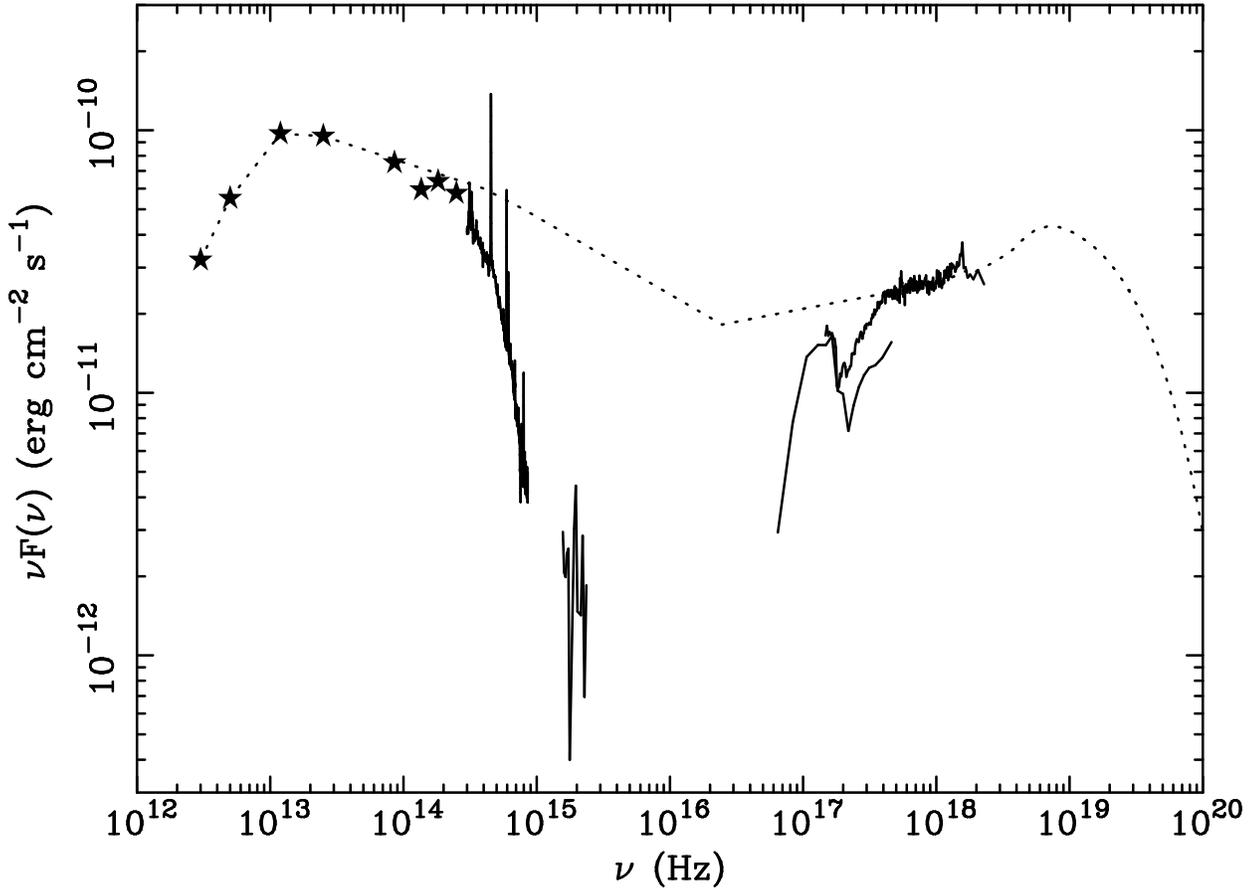,width=1.0\textwidth,angle=270}}
\caption{Multiband spectrum of MCG$-$6-30-15.   These data have not been
corrected for reddening (either Galactic or intrinsic) and include the
stellar component.  The dotted line shows the estimated intrinsic spectrum
of the source, after correction for reddening and absorption.  A reddening
of $E(B-V)=0.61$ (see Section 3) has been assumed in the placement of this
line.}
\end{figure*}

\section{Extinction and absorption}

The NIR/optical/UV spectrum of MCG$-$6-30-15 displays the signatures of
extinction and reddening by dust.  In this section, these signatures are
discussed and the amount of reddening is estimated.  We then compare the
amount of reddening with the observed X-ray absorption.

\subsection{Extinction of the optical emission line regions}

Extinction by cosmic dust is highly wavelength dependent and hence can
change observed line flux ratios significantly away from the intrinsic
(i.e. emitted) values.  This provides a classic method for determining the
amount of dust extinction along the line of sight to a particular emission
line region.  The relative ratios of the Balmer lines of hydrogen are often
used as extinction indicators due to the fact that they are observationally
convenient (being in the optical band), strong and their intrinsic relative
flux ratios are fairly well determined from atomic theory.

In the case of Balmer lines from AGN, one would ideally deblend the lines
into kinematically distinct components, e.g. a broad component (from the
BLR) and a narrow component (from the NLR).  One could then obtain
information about the extinction through to each component.  However,
high-resolution data is required to facilitate the deblending.  To avoid
introducing uncertainties due to the deblending procedure we choose to use
the total Balmer decrements: thus, the extinction estimates below should be
considered as average values over all of the emission line regions.

\begin{table*}
\begin{center}
\begin{tabular}{lccccc}\hline
Balmer & observed & intrinsic & $a$ & $E(B-V)$ & $N_{\rm H}$ \\
decrement & ratio ${\cal R}$ & ratio ${\cal R}_{\rm intr}$ & & &
($10^{20}\pcmsq$) \\\hline
H$\alpha$/H$\beta$ & 8.0 (7.6--8.4) & 2.76 & 2.21 & 1.02 (0.61--1.09) & 42
(26--44)\\
H$\gamma$/H$\beta$ & 0.31 (0.22--0.37) & 0.474 & $-5.17$ & 0.95
(0.56--1.72) & 39 (24--69) \\
H$\delta$/H$\beta$ & 0.16 (0.10--0.26) & 0.262 & $-3.52$ & 0.75
(0--1.47) & 31 (0--60)\\\hline
\end{tabular}
\caption{Balmer decrements and inferred extinction for MCG$-$6-30-15.  See
text for a discussion of the significance and possible causes of
differences between $E(B-V)$ as calculated from difference Balmer
decrements.  The figures in brackets are the allowed range of the parameter
given the various systematic effects discussed in the text.}
\end{center}
\end{table*}

Table~2 gives the observed Balmer decrements and the expected intrinsic
value based upon the assumption of case-B recombination.  These decrements
have been converted into the reddening, $E(B-V)$, using the standard
interstellar extinction curve of Osterbrock (1989).  This interstellar
extinction curve leads to the expression
\begin{equation}
E(B-V)=a\log\left(\frac{{\cal R}}{{\cal R}_{\rm intr}}\right)
\end{equation}
where ${\cal R}$ is the observed Balmer decrement, ${\cal R}_{\rm intr}$ is
the intrinsic Balmer decrement and $a$ is a constant which is given in
Table~2 for the three Balmer decrements quoted.  Table~2 also associates a
hydrogen column density, $N_{\rm H}$, with this reddening.  This is given
by (Heiles, Kulkani \& Stark 1981)
\begin{equation}
N_{\rm H}=(2.14\pm 0.20)\times 10^{20} + (3.91\pm 0.21)\times 10^{21}\,E(B-V)\,\,\pcmsq
\end{equation}
and is the column density of cold gas that would be present under the
assumption that the nature of the dust and the cold-gas/dust ratio is the
same as is found locally in our Galaxy.

The calculation of the uncertainties in $E(B-V)$ deserves discussion.  We
must critically consider both the observational uncertainties and the
intrinsic nature of the source.

First, we consider observational uncertainties.  The H$\alpha$ line and
H$\beta$ line were measured in different spectrographs, the data from which
have undergone independent reduction and calibration.  This will introduce
some uncertainty into the H$\alpha$/H$\beta$ Balmer decrement: in Section
2.1 we estimated that both the red and blue spectra have absolute
normalizations which are uncertain to $\sim 3$ per cent (with these two
uncertainties being independent).  This leads to a 4--5 per cent in the
H$\alpha$/H$\beta$ Balmer decrement, corresponding to an error on the
reddening of $E(B-V)_{H\alpha/H\beta}=1.02\pm 0.05$.  The H$\beta$,
H$\gamma$ and H$\delta$ lines all appear in the data from the same spectrograph
and so the two independent Balmer decrements that can be formed from these
three lines are not sensitive to uncertainties in the overall
normalization.  However, H$\gamma$ and H$\delta$ do suffer from potentially
significant observationally-related uncertainties.  The H$\gamma$ line is
blended with the [OIII]$\lambda 4363$ line and so an unambiguous
determination of the H$\gamma$ flux, $F(H\gamma)$, is difficult without
higher resolution data.  By examining extreme cases, we can conservatively
bracket $F(H\gamma)$ to the range $(2.8-4.3)\times 10^{-14}\ergpcmsqps$.
The corresponding range of inferred reddening is
$E(B-V)_{H\gamma/H\beta}=$0.56--1.72.  The H$\delta$ flux, $F(H\delta)$, is
uncertain due to the fact that it is weak compared with the local continuum
and so is sensitive to the modeling of that continuum.  By examining
extreme cases, we can conservatively bracket $F(H\delta)$ to $(1.3-3.0)\times
10^{-14}\ergpcmsqps$.  The corresponding range of inferred extinctions is
$E(B-V)_{H\delta/H\beta}=$0--1.47.

Secondly, we must consider the possibility that the intrinsic Balmer
decrements are not well represented by their case-B recombination limits.
Deviations from case-B values can occur due to collisional effects and
radiative transfer effects which are especially important in the
high-density gas found within the BLR.  Theoretically, these processes can
increase the H$\alpha$/H$\beta$ decrement from the case-B value to 10 (Kwan
\& Krolik 1981) or more (Canfield \& Puetter 1981).  In these extreme
cases, we would not have to postulate any dust reddening towards the
emission line regions of MCG$-$6-30-15 at all.  However, observations of
other Seyfert 1 nuclei which are thought to be unreddened suggest that
intrinsic H$\alpha$/H$\beta$ decrements do not exceed 4 (e.g. Malkan 1983;
Wu, Boggess \& Gull 1983).  Using this value of the H$\alpha$/H$\beta$
ratio instead of the case-B value decreases the inferred reddening from
$E(B-V)=1.02$ to $E(B-V)=0.67$.  Including the uncertainty in the measured
H$\alpha$/H$\beta$ ratio, the lower limit to the reddening is
$E(B-V)=0.61$.  The effect of deviations from case-B recombination on the
intrinsic H$\gamma$/H$\beta$ and H$\delta$/H$\beta$ have not been
investigated in as much theoretical detail as for H$\alpha$/H$\beta$.
However, the corresponding uncertainties on the reddening are likely to be
insignificant compared with the observational uncertainties discussed
above.

To summarize these Balmer decrement studies, the large H$\alpha$/H$\beta$
ratio suggests a reddening in the range $E(B-V)=0.61$--1.09.  This is
compatible with the reddening inferred from H$\gamma$/H$\beta$ and
H$\delta$/H$\beta$ although these two Balmer decrements have large
uncertainties due to line blending and the modeling of the
continuum. Assuming that the cold-dust/gas ratio is similar to that
observed locally in our Galaxy, the column density of gas associated with
this reddening is in the range $N_{\rm H}=(2.6-4.4)\times
10^{21}\pcmsq$.

\subsection{Extinction of the continuum source}

The optical continuum radiation will also be affected by extinction and
reddening.  Suppose the {\it intrinsic} spectrum of the (optical) continuum
has a form $F_{\rm intr}(\nu)\propto\nu^{-\alpha}$.  Furthermore, suppose
that the {\it observed} (i.e. reddened) optical continuum spectrum is
$F_{\rm obs}(\nu)\propto \nu^{-\alpha_{\rm obs}}$.  The difference between
$\alpha$ and $\alpha_{\rm obs}$ is related to the reddening towards the
source.  From the definition of reddening, it is easily seen that
\begin{eqnarray}
E(B-V)&=&-2.5\log\left(\frac{\nu_{\rm B}^{-\alpha_{\rm obs}}}{\nu_{\rm V}^{-\alpha_{\rm obs}}}\right)+2.5\log\left(\frac{\nu_{\rm B}^{-\alpha}}{\nu_{\rm V}^{-\alpha}}\right)\\
&=&2.5(\alpha_{\rm obs}-\alpha)\log\left(\frac{\nu_{\rm B}}{\nu_{\rm
V}}\right)\\
&\approx &0.26(\alpha_{\rm obs}-\alpha)
\end{eqnarray}
where $\nu_{\rm B}$ and $\nu_{\rm V}$ are the frequencies characterizing the
$B$-band and $V$-band respectively.  

Unreddened Seyfert 1 nuclei in the luminosity range occupied by
MCG$-$6-30-15 tend to have optical continua with $\alpha\sim 2$ (e.g. see
study of Morris \& Ward 1988, and note that a flat flux spectrum in
wavelength space implies $\alpha=2$ when considered in frequency space).
Fitting the (galaxy-subtracted) blue spectrum of the nucleus of
MCG$-$6-30-15, we conclude that $\alpha_{\rm obs}$ lies in the range
4.5--5.  Assuming that the intrinsic optical spectrum of MCG$-$6-30-15 is
similar to that found in unreddened Seyfert 1 nuclei, the reddening of the
optical continuum source $E(B-V)$ is in the range 0.65--0.78.

The reddening of the optical continuum source is consistent with the lower
end of the reddening derived from the H$\alpha$/H$\beta$ Balmer ratio.
Assuming that the optical continuum is associated with an accretion disk
embedded inside the BLR, this result suggests that little dust is present
between the accretion disk and the region where the bulk of the the broad
line photons are emitted.

\subsection{The UV flux -- evidence for scattering?}

Initially, we shall suppose that the observed {\sc C\,iv} line emitting
region is seen directly rather than via scattered photons.  Furthermore,
suppose that the UV continuum and {\sc C\,iv} line emission are subject to
the same extinction as the optical non-stellar continuum/line regions.
Since we have determined the optical reddening to be in the range
$E(B-V)=0.61-1.09$, we can deredden the {\sc C\,iv} line in order to
estimate its intrinsic (i.e. dereddened) flux.  Thus, {\it given this
supposition that the UV line emitting region is seen directly}, we
constrain the intrinsic {\sc C\,iv} line flux to lie in the range
\begin{equation}
F_{\rm intr}(CIV)=(1.4-102)\times 10^{-11}\ergpcmsqps,
\end{equation}
where we have included the 1-$\sigma$ errors in the observed flux and used
the UV extinction law of Osterbrock (1989).  This is a rather large line
flux, corresponding to an isotropic luminosity of $3\times 10^{42}\ergps$
or greater in the {\sc C\,iv} line alone.

To quantitatively assess how large this line flux is, consider the lower
end of this range corresponding to $E(B-V)=0.61$.  For this reddening, the
dereddened H$\beta$ flux is
\begin{equation}
F_{\rm intr}(H\beta)=9.1\times 10^{-13}\ergpcmsqps,
\end{equation}
leading to a lower limit on the intrinsic {\sc C\,iv}/H$\beta$ flux ratio
of 15.  This ratio is very sensitive to the reddening assumed and can {\it
greatly} exceed this value if $E(B-V)>0.61$.  In unreddened Seyfert nuclei,
this ratio is often significantly smaller.  For example, in the AGN Watch
Campaign on NGC~3783, the intrinsic {\sc C\,iv}/H$\beta$ flux ratio is
$\sim 10$ (Reichert et al. 1994; Stirpe et al. 1994).  Similarly, the {\sc
C\,iv}/H$\beta$ flux ratios found during the monitoring campaigns on
NGC~4151 (Crenshaw et al. 1996; Kaspi et al. 1996) and NGC~5548 (Korista et
al. 1995) are $\sim 7$ and $\sim 9$, respectively.

We must conclude that the {\sc C\,iv} line flux is unusually high compared
with the optical line fluxes, or that one of our assumptions has broken
down.  There are three possible ways that our above argument might be
flawed.  First, source variability during the 9 months separating the UV
and optical observations may produce an apparently unusual line ratio, even
if the intrinsic line ratio is normal.  In our minimum reddening case
($E(B-V)=0.61$), only mild variability ($\sim 30$ per cent over 9 months)
is required to make the observed {\sc C\,vi}/H$\beta$ ratio of 15
consistent with the that seen in other objects.  As one postulates higher
reddening values, the more extreme is the inferred intrinsic line ratio and
the more violent the variability needed.  Secondly, the reddening towards
the high-ionization BLR (including the {\sc C\,iv} line emitting region)
may be different than that towards the low-ionization BLR (which includes
the Balmer line emitting region).  Whilst this is clearly a viable
possibility (and one can imagine central-engine geometries that produce
such an effect) there is no precedent for the high-ionization BLR to be
less reddened than the low-ionization BLR.  Thirdly, some fraction of the
photons from the BLR might be scattered around the material responsible for
the extinction.  If the scattering fraction is wavelength independent
(e.g. electron scattering), the scattering will tend to preferentially
enhance the UV relative to the optical due to the fact that the {\it
direct} flux is heavily reddened.  Since we know scattering to be an
important process in some other Seyfert nuclei, we now explore this last
possibility in more detail.

Suppose that the intrinsic UV/optical line spectrum is similar to that of
NGC~3783, with a {\sc C\,iv}/H$\beta$ flux ratio of 10.  We can write the
observed fluxes of both of these lines, $F_{\rm obs}$, as a sum of the
direct (extinguished) flux and the scattered flux which is assumed not to
suffer any extinction beyond that due to Galactic material.  If $f$ is the
scattering fraction, then we have
\begin{equation}
F_{\rm obs}=\left(10^{-b E_{\rm Gal}(B-V)}f+10^{-b E(B-V)}\right)F_{\rm intr}
\end{equation}
where $b$ is a parameter dependent on the extinction law used.  The
standard interstellar extinction curve of Osterbrock (1989) gives $b=3.2$
for {\sc C\,iv}$\lambda 1549$ and $b=1.45$ for H$\beta$.  The first term on
the right hand side of equation (8) represents the scattered flux including
the effects of extinction by Galactic material.  We take $E_{\rm
Gal}(B-V)=0.06$ (Berriman 1989).  The second term of equation (8) gives the
contribution due to the direct (extinguished) flux.  Dividing these
equations for {\sc C\,iv}$\lambda 1549$ and H$\beta$ gives a relation
between the required scattering fraction $f$ and the total line-of-sight
reddening $E(B-V)$.  This relationship is shown in Fig.~5 for interesting
values of $E(B-V)$.  It can be seen that scattering fraction of between
1--5 per cent (depending on the total reddening) are required in order to
make the observed line ratios consistent with an intrinsic {\sc
C\,iv}/H$\beta$ line ratio of 10.

\begin{figure}
\hbox{
\hspace{1cm}
\centerline{\psfig{figure=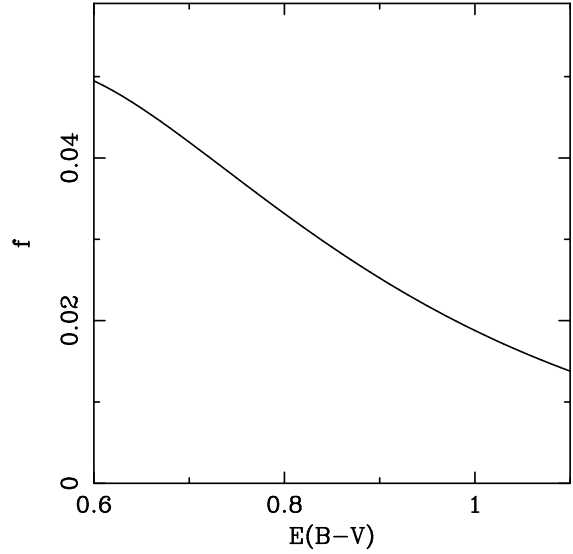,width=0.65\textwidth,angle=270}}
}
\caption{Relationship between the required scattering fraction, $f$, and the
total line-of-sight reddening, $E(B-V)$, assuming an intrinsic {\sc
C\,iv}$\lambda 1549$/H$\beta$ flux ratio of 10.}
\end{figure}

\subsection{X-ray absorption}

X-rays are thought to be produced in the very central regions of the
accretion disk.  The absorption of those X-rays by line-of-sight material
is readily observable at soft X-ray energies and provides important
information on the surroundings of the active nucleus.  The discussion of
X-ray absorption is commonly divided into that of `cold' absorption
(i.e. by neutral material) and `warm' absorption (i.e. by ionized
material). 

\begin{table}
\begin{center}
\begin{tabular}{lll}\hline
property & inner absorber & outer absorber \\\hline
$N_{\rm W}$ $(10^{20}\pcmsq)$ & 130 & 46 \\
$\xi$ $(\erg\cm\ps)$ & 74 & 17 \\
density $n_{\rm e}$ $(10^5\pcmcu)$ & $\approxgt 20$ & $\approxlt 2$\\
distance $R$ ($\pc$) & $\approx 0.005$ & $\approxgt 1$ \\
temperature $T$ $(10^5\K)$ & 5 & $0.5$ \\
$[{\rm FeX}]_{\rm model}$/$[{\rm FeX}]_{\rm obs}$ & $4\times 10^{-5}f_{\rm c}$ & $24f_{\rm c}$\\
$[{\rm FeXI}]_{\rm model}$/$[{\rm FeXI}]_{\rm obs}$ & $4\times 10^{-6}f_{\rm c}$ & $0.5f_{\rm c}$\\
$[{\rm FeXIV}]_{\rm model}$/$[{\rm FeXIV}]_{\rm obs}$ & $0.04f_{\rm c}$ & $24f_{\rm c}$ \\\hline
\end{tabular}
\caption{Properties of the warm absorbers in MCG$-$6-30-15.}
\end{center}
\end{table}

The warm absorption has already been discussed in Sections 2.3.2 and 2.3.3.
As previously mentioned, variability studies have led to a two-zone model
for this absorber.  Table~3 summarizes the properties of these two
absorbing regions based on modeling with the photoionization code {\sc
cloudy} (Ferland 1991).

Once the effect of the warm absorber has been modeled, the X-ray spectrum
can be examined for evidence of additional cold absorption in excess to
that expected from our own Galaxy (the Galactic column density along the
line of sight to this source is $N_{\rm Gal}=4.1\times 10^{20}\pcmsq$).
Describing the warm absorber with a simple one-zone photoionization model
computed by the code {\sc cloudy} (see Fabian et al. 1994 for a detailed
description of this warm absorber model), the {\it ASCA} spectrum suggests
an excess cold column density of $N_{\rm H}=1.7^{+0.4}_{-0.3}\times
10^{20}\pcmsq$.  Quoted errors are statistical in nature and stated at the
90 per cent confidence level for one interesting parameter
($\Delta\chi^2=2.7$).  It has been suggested that the low energy
calibration of the {\it ASCA} SIS is incorrect so as to over-estimate the
cold absorption by 1--3$\times 10^{20}\pcmsq$ (e.g. see discussion in Cappi
et al. 1997).  Thus, within these calibration uncertainties, our result may
be consistent with there being negligible cold absorption excess to the
Galactic column.  Fitting the same simple model to the {\it ROSAT} PSPC
spectrum suggests an excess column density of $N_{\rm H}=(1.9\pm 0.3)\times
10^{20}\pcmsq$, in good agreement with the {\it ASCA} result.

\subsection{The absorption/reddening mis-match}

Comparing the results of Sections 3.1--3.4 reveals an apparent
contradiction.  X-ray observations show that the column density of cold
(neutral) gas along the line of sight to the primary X-ray source is
$N_{\rm H}\approxlt 2\times 10^{20}\pcmsq$.  However, both the optical
continuum source and the BLR are highly reddened.  On the basis of local
(Galactic) studies, we would expect a cold gas column density of $N_{\rm
H}\approxgt 2.6\times 10^{21}\pcmsq$ to be associated with the dust
responsible for the reddening.  This is more than an order of magnitude
above the X-ray limits.  Since the X-ray emission is thought to occur
deeper within the central engine than either the optical continuum or
optical line emission, this result is somewhat surprising.

This discrepancy for MCG$-$6-30-15 was first hinted at by Pineda et
al. (1980) and explicitly commented upon by Reynolds \& Fabian (1995) on
the basis of data drawn from the literature.  A similar discrepancy has
been found for IRAS~13349$+$2438 (Brandt, Fabian \& Pounds 1996), an
infrared luminous quasar which also displays a prominent warm absorber.
These authors discuss various resolutions of this discrepancy.  To
summarize these discussions, it is found that the only way to reconcile
this result with a plausible geometry and a physically reasonable
gas-to-dust ratio is to suggest that the dust resides in the {\it ionized}
gas that constitutes the warm absorber.  The soft X-ray opacity of this
material is less than that of cold material primarily due to the almost
complete ionization of hydrogen and helium.  The dusty warm absorber
hypothesis has also been explored by Reynolds (1997) in the context of a
sample of Seyfert galaxies.  We now discuss dusty warm absorbers in more
detail.

\section{Dusty warm absorbers}

In this section we explore the idea of a dusty warm absorber in more detail
than any of the above previous work.  In particular, we construct
photoionization models of dusty warm absorbers and explicitly fit these
models to the {\it ASCA} data.  

\subsection{Dust survivability}

Dust grains are highly efficient radiators and hence can thermally decouple
from the surrounding hot gas.  Under the conditions envisaged here, there
are two grain destruction mechanisms that must be considered.  First, if
the grains themselves become too hot, they will rapidly sublime.  The grain
temperature will be set by the thermal equilibrium between the AGN
radiation incident on a given grain and the thermal radiation emitted by
that grain (e.g. see Barvainis 1987).  For MCG$-$6-30-15, the {\it
sublimation radius} (i.e. the radius from the AGN within which dust
grains become so hot that they sublime) is $\sim 10^{17}\cm$.  Thus, any
dust grains within the inner warm absorber would be rapidly sublimed by the
intense radiation field.  Dust in the outer warm absorber would not be
subject to significant sublimation.

The second dust destruction mechanism that we must consider is thermal
sputtering.  If we make the standard assumption that the (outer) warm
absorber is photoionized, then photoionization models suggest that the gas
temperature is only $T\sim 5\times 10^4\K$ and thermal sputtering is
negligible.  If, instead, we suppose that the outer warm absorber is purely
collisionally-ionized, gas temperatures of $T\sim 10^6\K$ are required in
order to achieve the observed ionization states (Shull \& van Steenberg
1982).  From the expressions of Burke \& Silk (1974), the thermal
sputtering timescale for this temperature is
\begin{equation}
t_{\rm sp}\approx 3\times 10^6 \left(\frac{n}{1\pcmcu}\right)^{-1}\yr,
\end{equation}
where $n$ is the electron number density in the gas.  Suppose that $r$ is
the distance of the outer warm absorber from the central engine, and $L$ is
the (ionizing) luminosity of the central engine.  Furthermore, define
$\xi_{\rm equiv}\approx 20\erg\cm\ps$ to be the ionization parameter of a
photoionized plasma in which oxygen is ionized to the same degree as seen
in the outer warm absorber of MCG$-$6-30-15.  Given our (temporary)
hypothesis that the plasma is collisionally-ionized, the density must
satisfy
\begin{equation}
n>\frac{L}{\xi_{\rm equiv} r^2},
\end{equation}
or else photoionization would dominate the ionization state.   Evaluating
the sputtering timescale for the parameters of MCG$-$6-30-15 gives
\begin{equation}
t_{\rm sp}\approxlt 10\,\left(\frac{r}{1\pc}\right)^2\yr.
\end{equation}
For comparison, the flow timescale of the outer warm absorber is
\begin{equation}
t_{\rm flow}\sim \frac{r}{v_{\rm flow}}\sim 10^3\left(\frac{r}{1\pc}\right)\yr,
\end{equation}
where we have adopted a typical value of $v_{\rm flow}=1000\kmps$ for the
velocity of the outer warm absorber, as indicated by UV absorption line
studies of other AGN (Mathur, Elvis \& Wilkes 1995).  It can be seen that
the flow timescale of the warm absorber always exceeds the sputtering
timescale unless $r\approxgt 100\pc$.  If the outer warm absorber was
situated at such a large distance, then either we would have to be viewing
the AGN along a very special line of sight, or else the mass, $M$, and
kinetic energy, $L_{\rm K}$, associated with the outflow would both be
huge.  From the expressions of Reynolds \& Fabian (1995), and assuming a
global covering fraction of $f_{\rm c}=0.1$, we get $M\sim 10^6\Msun$ and
$L_{\rm K}\sim 3\times 10^{42}\ergps$.  The initial acceleration of this
material would be extremely problematic to understand.  We consider this
possibility to be unphysical.  Thus, in the absence of a viable,
collisionally-ionized model, {\it the observation of a dusty warm absorber
may be taken as further evidence that photoionization dominates the state
of this plasma.}

Whilst dust can survive in warm photoionized gas, it is extremely
difficult to form dust in such an environment: the grains could never
assemble at such temperatures.  Furthermore, a comparison of the column
density of the warm absorber with the cold column expected to be associated
with the reddening reveals that the warm-gas/dust ratio in the warm
absorber must be very similar to the cold-gas/dust ratio in our Galaxy.
These two facts taken together suggest that the warm material originates
from dusty cold material, possibility via radiative heating, and that a
substantial fraction of the dust survives the heating process.  The
putative dusty molecular torus of Seyfert unification schemes might be a
possible progenitor of such a radiatively-driven, warm, dusty outflow.

\subsection{Photoionization models}

We have constructed photoionization models of dusty warm absorbers using
the photoionization code {\sc cloudy}.  Grids of such models were
constructed for various values of the column density $N_{\rm W}$,
ionization parameter $\xi$ and X-ray photon indices $\Gamma$.  Since we are
interested in the behaviour of the outer warm absorber, the distance of the
absorber from the primary source was fixed at 1\,pc.  Otherwise, these
models are identical to those of Fabian et al. (1994) and Reynolds et
al. (1995) except for the inclusion of dust grains.  The grain models used
are described in {\sc hazy} (the manual to {\sc cloudy}) pp.~284.
Prompted by the observations of the previous paragraph, we have fixed the
gas/dust ratio to Galactic value.  Two such grids were computed: one
contains a standard (i.e. local) mixture of silicate and graphite grains
whereas the other contains only graphite grains.

\begin{table}
\begin{center}
\begin{tabular}{lcc}\hline
model & standard & graphite \\
parameter & dust mixture & dust only \\\hline
$N_{\rm W,outer} (10^{20}\pcmsq)$ & $47.1^{+2.9}_{-2.0}$ & $57.5^{+5.6}_{-3.8}$\\
$\xi (\erg\cm\ps)$ & $25.1^{+3.8}_{-0.3}$ & $19.5^{+2.6}_{-2.3}$ \\
$\Gamma$ & $2.17^{+0.02}_{-0.04}$ & $2.13^{+0.02}_{-0.02}$ \\
$\chi^2$/dof & 430/229 & 362/229 \\\hline
\end{tabular}
\end{center}
\caption{Results of fitting dusty warm absorber models to {\it ASCA} data.}
\end{table}

These models were fitted to the {\it ASCA} data.  Since we are interested
in modeling fine details of the soft {\it ASCA} spectrum, only data from
the best calibrated solid-state imaging spectrometer (SIS0) were used in
the spectral fitting process.  Furthermore, only data in the range
0.6--4\,keV were fitted: below 0.6\,keV the {\it ASCA} calibration becomes
uncertain whereas above 4\,keV spectral complexities due to the Fe
K$\alpha$ line become relevant.  In detail, our spectral model has three
components.  First, the primary power-law and the {\it outer} warm absorber
are modeled using {\sc cloudy} as described above.  Secondly, the effect
of the {\it inner} warm absorber was modeled as a {\sc O\,viii} absorption
edge with (rest-frame) threshold energy 0.87\,keV and optical depth at
threshold of $\tau_{\rm O8}=0.18$ (Otani et al. 1996)\footnote{Short term
spectral variability in the soft {\it ASCA} band is known to occur during
this observation (Otani et al. 1996), and one might worry about the effect
of this variability when one performs spectral fitting on the time-averaged
spectrum.  However, it is known that changes in $\tau_{\rm O8}$ dominate
this spectral variability.  Thus, the best fitting parameters for the outer
warm absorber (which contributes very little to the {\sc O\,viii} edge)
should be relatively unaffected by spectral variability}.  Lastly, Galactic
absorption by a cold column of $N_{\rm H}=4.06\times 10^{20}\pcmsq$ was
included.  The spectral fitting results are shown in Table~4.

\begin{figure*}
\hbox{
\psfig{figure=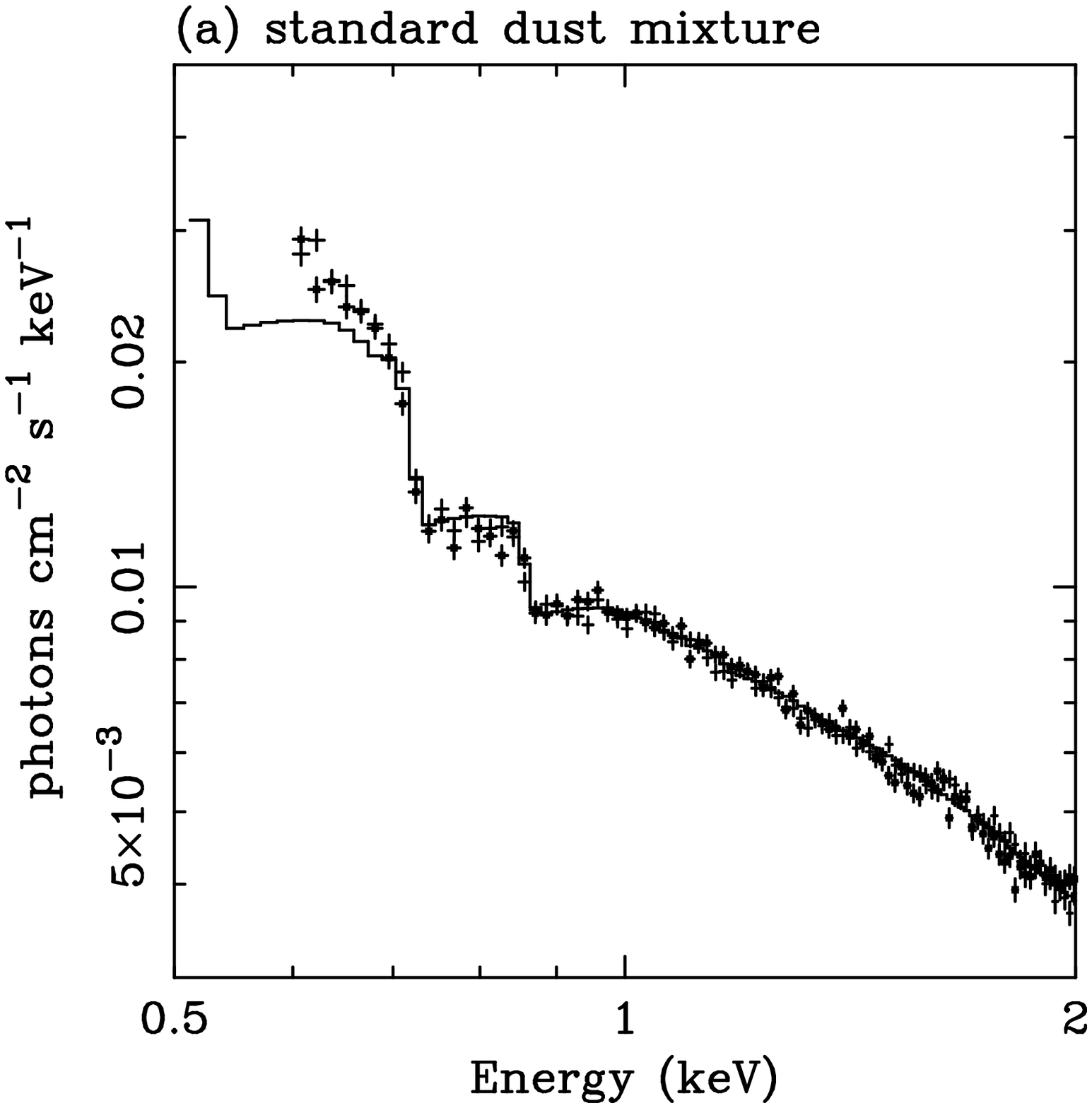,width=0.6\textwidth,angle=270}
\hspace{-2cm}
\psfig{figure=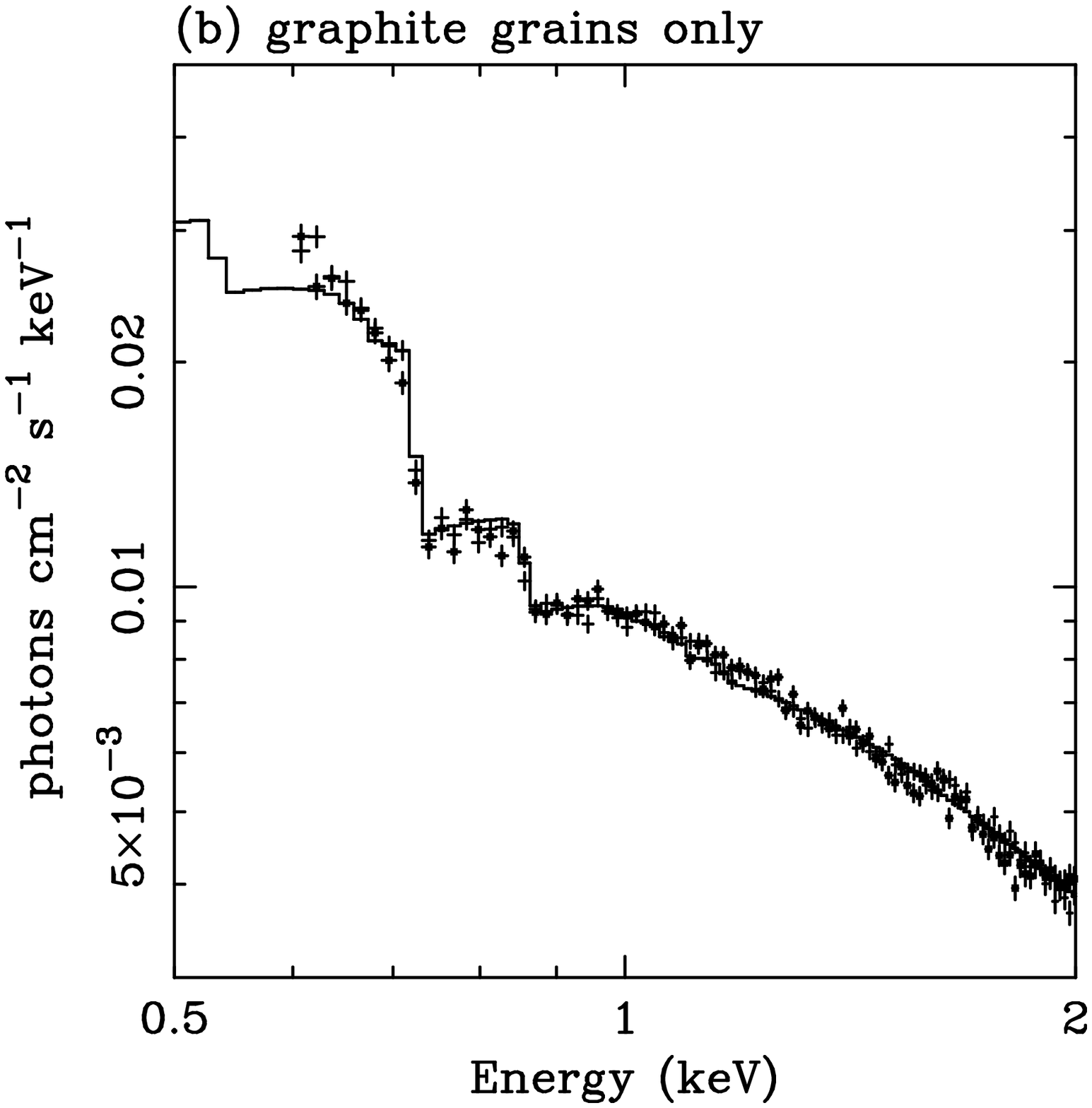,width=0.6\textwidth,angle=270}
}
\caption{Dusty warm absorber models fitted to 0.6--4\,keV data from the
{\it ASCA} SIS0 (plain crosses).  SIS1 data are also shown (filled squares)
but have not been used in the spectral fitting reported in Table~4.  Panel
a) shows the model computed with a standard (i.e. Galactic) mixture of
graphite and silicate dust grains.  Panel b) shows the model computed with
only graphite grains.}
\end{figure*}

Table~4 shows that the assumed dust composition has a significant effect on
the goodness of fit.  The model which assumes graphite grains only is a
much better fit than the model with a standard dust mixture
($\Delta\chi^2=68$ for the same number of dof).  The reason for this
difference is illustrated in Fig.~6 which shows the best-fit unfolded model
and the 0.6--2\,keV SIS0 data for each of the two assumed dust
compositions.  The standard dust model predicts a large photoelectric
K-edge due to neutral oxygen (threshold energy $0.533\keV$).  Such an edge
is not observed. The graphite grain model predicts a significantly smaller
neutral oxygen edge which is much more consistent with observations.  Note
that the neutral oxygen edge in the latter model originates purely from the
Galactic column and not from the dusty warm absorber.  Thus, the graphite
grain model seems to be preferred over the standard dust mixture model.

The signatures of dust in a dusty warm absorber only become significant at
X-ray energies below the {\it ASCA} band.  Thus, we must check that the
dusty warm absorber model is consistent with the soft X-ray spectrum as
determined by the {\it ROSAT} PSPC.  In detail, we compared the {\it ROSAT}
PSPC data with a spectral model consisting of a power-law form (photon
index $\Gamma$) absorbed by three components:

a) a dusty warm absorber model as computed by {\sc cloudy} (column density
$N_{\rm W}$ and ionization parameter $\xi$), 

b) an absorption edge at the threshold energy of {\sc O\,viii} to mimic the
effect of the dust-free inner warm absorber (optical depth at threshold
$\tau_{\rm O8}$),

c) neutral absorption to account for Galactic and intrinsic cold gas
absorption (column density $N_{\rm H}$).

This 6 parameter scheme over-models the {\it ROSAT} PSPC spectrum (which
has $\sim 5$ independent energy channels).  Thus, we do not formally fit
the data since any such fit is very poorly constrained -- we merely seek to
demonstrate consistency with the fit parameters derived from the {\it ASCA}
data.  Both the standard dust and the graphite dust warm absorber models
are found to be consistent with the PSPC data for the following parameters:
$\Gamma=2.8$, $N_{\rm W}=4.5\times 10^{21}\pcmsq$, $\xi=20\erg\cm\ps$,
$\tau_{\rm O8}=0.83$, $N_{\rm H}=6.9\times 10^{20}\pcmsq$.  These
parameters are roughly consistent with those derived from the {\it ASCA}
data with the exception of $\Gamma$ and $\tau_{\rm O8}$.  We will briefly
address these in turn.

First, the photon index $\Gamma$ is inferred to be significantly steeper in
the {\it ROSAT} observation than the {\it ASCA} observation.  At least some
of this discrepancy ($\Delta\Gamma\sim 0.5$) may be due to previously noted
errors in the {\it ROSAT}--{\it ASCA} cross-calibration.  However, there
may be a true softening of the X-ray spectrum in the {\it ROSAT} band due
to the onset of a soft excess.  It must be noted that the {\it ASCA} data
do not show any evidence of a soft excess above $0.6\keV$ (Reynolds 1997).
Temporal variations of $\Gamma$ may also explain such a discrepancy (note
that the {\it ROSAT} and {\it ASCA} observations are separated by over 2
years).

Secondly, the {\sc O\,viii} edge depth $\tau_{\rm O8}$ is inferred to be
significantly deeper in the {\it ROSAT} observation than in {\it ASCA}
observation.  This can be understood as a real (i.e., physical) change.  Otani
et al. (1996) and Reynolds (1996) have found a relationship between the
instantaneous value of $\tau_{\rm O8}$ (as measured by {\it ASCA}) and the
luminosity of this source.  From the analysis of Reynolds (1996), this
relation takes the form
\begin{equation}
\tau_{\rm O8}\approx 0.12\left(\frac{L_{2-10}}{10^{43}\ergps}\right)^{-2.7}
\end{equation}
where $L_{2-10}$ is the 2--10\,keV luminosity of the source. This can be
understood physically in terms of a highly-ionized warm absorber in which most
of the oxygen atoms are fully stripped of all electrons (i.e. {\sc O\,ix}
is the dominant state).  A drop in ionizing luminosity results in an
increased number of {\sc O\,viii} ions due to recombination of the {\sc
O\,ix} ions.  This produces the observed anti-correlation between
$\tau_{\rm O8}$ and $L_{2-10}$.  

During the {\it ROSAT} observation, the average 2--10\,keV luminosity is in
the range $L_{2-10}=4-8\times 10^{42}\ergps$.  The large uncertainty in
luminosity is due to the uncertainty in the extrapolation from the {\it
ROSAT} band to the 2--10\,keV band.  The corresponding range of edge depth
is $\tau_{\rm O8}=0.2-1.4$.  Thus, the value needed to agree with the {\it
ROSAT} spectrum, $\tau_{\rm O8}=0.83$, is completely consistent with this
relationship.

To summarize these X-ray results, we have shown that a warm absorber
containing sufficient dust to explain the optical reddening is also
compatible with the {\it ASCA} and {\it ROSAT} data.  In principle, a
detailed examination of the neutral K-edges of the various dust-phase
metals allows the composition of the dust to be probed.  Although it is
extremely hard to make definitive statements yet due to the lack of
high-quality soft X-ray spectra, there is evidence that the dust grain
composition is non-standard in so far as it contains few silicate grains.
We note that due to the tentative nature of this conclusion, we have not
taken account of any non-standard dust composition when performing the
reddening calculations of Section 3.  Clearly, this should be the subject
of future work.

\subsection{Dust emission and the infrared bump}

The multiwaveband spectrum of MCG$-$6-30-15 clearly shows a large IR bump.
If this infrared bump is interpreted as the thermal emission from dust
grains, the corresponding grain temperature is $T\sim 300-500\K$.  The fact
that the $25-100\mu {\rm m}$ fluxes describe an approximate Rayleigh-Jeans
form shows that dust cooler than $T\sim 300\K$ cannot contribute much to
the observed emission.  However, the relatively flat $12-25\mu {\rm m}$
spectrum shows that there are probably hotter dust components.  Here we
investigate whether this IR bump can be understood as thermal emission from
a dusty warm absorber.

Following the work of Barvainis (1987), we calculate the mass of dust
required to produce the $12\mu {\rm m}$ flux observed in MCG$-$6-30-15.
Assuming graphite grains with a radius of $a=0.05\mu {\rm m}$, a density of
$\rho=2.26\g\pcmcu$ and temperature of $T=500\K$, we estimate that
$15\Msun$ of dust is required to produce the $12\mu {\rm m}$ flux.
Furthermore, assuming a standard gas-to-dust mass ratio of 200, the
corresponding mass of associated gas is $3000\Msun$.  If, instead, we
consider a grain temperature of $T=1500\K$, the increased dust emissivity
lowers the dust mass requirement to $6\Msun$ with a corresponding gas mass
of $1200\Msun$.

The mass of the outer (dusty) warm absorber can be independently
estimated from simple arguments using the X-ray data.  Approximating as a
thin, uniform, spherical shell, the mass of ionized plasma is
\begin{eqnarray}
M&=&4\pi R^2 N m_{\rm p} f_{\rm c}\\
&\approx &1000\,R_{\rm pc}^2 N_{22} f_{\rm c}\,\Msun
\end{eqnarray}
where $R=R_{\rm pc}\pc$ is the radial distance of the absorbing region,
$N=10^{22}N_{22}\pcmsq$ is the column density and $f_{\rm c}$ is the
covering fraction of the material as seen from the central source (i.e. the
`global' covering fraction).  In MCG$-$6-30-15, we must have $R_{\rm
pc}\approxlt 1$ in order to heat the dust to $T\approxgt 500\K$.  The {\it
ASCA} data directly constrain the column density to be $N_{22}\sim 0.5$.
Thus, we have
\begin{equation}
M\approxlt 500\,f_{\rm c}\,\Msun
\end{equation}
with only $f_{\rm c}$ remaining unconstrained.  This is somewhat less than
the mass derived in the previous paragraph, especially when it is noted
that there are arguments suggesting $f_{\rm c}\sim 0.1$ (see Reynolds 1997
and Section 5.)

The disagreement between these two mass estimates is not unexpected.  We
would only expect agreement if most of the warm/hot dust emission
originated from grains in a warm absorber which approximated a uniform
spherical shell.  Realistically, a rather more complicated geometry would
be expected.  In particular, large quantities of warm/hot grain emission
would be expected from dusty gas associated with the irradiated inner edge
of the putative dusty molecular torus.  Indeed, it may be artificial to
consider the outer warm absorber and the torus as distinct and separate
entities -- the outer warm absorber may well be an optically-thin,
outflowing extension of the standard (optically-thick) cold torus.

\section{High-ionization (coronal) emission lines}

The optical spectrum of MCG$-$6-30-15 clearly displays the very
high-excitation forbidden lines of [FeX]$\lambda 6375$, [FeXI]$\lambda
7892$ and [FeXIV]$\lambda 5303$.  These are the so-called iron coronal
lines.  Previous studies of the coronal lines in Seyfert galaxies have
found them to have line widths intermediate between that of the
lower-excitation forbidden narrow lines (such as [OIII]) and the permitted
broad lines (Grandi 1978).  These studies have also found the coronal lines
to be slightly blueshifted with respect to the narrow forbidden lines.  As
shown in Table~2, the coronal lines in MCG$-$6-30-15 follow exactly these
trends.  Note that the true width and flux of the [FeXIV]$\lambda 5303$
line may be less than that presented in Table~2 due to blending with
[CaV]$\lambda 5309$.  Oliva et al. (1994) have examined various models for
the coronal line emission. 

Such observations have led to the discussion of the coronal line region
(CLR) which is distinct from the BLR or NLR.  Here we address the
possibility that the CLR in MCG$-$6-30-15 can be identified with the outer
warm absorber.

The collision strengths for these transitions are poorly known (Osterbrock
\& Fulbright 1996; Oliva 1996).  Any calculations of line strengths that we
perform will be tainted by this basic uncertainty in the atomic physics.
Despite these uncertainties, we have used {\sc cloudy} to examine the
coronal line emission from the inner and outer warm absorber.
Table~3 reports the predicted coronal line fluxes as a fraction of the
observed line flux.  These predicted fluxes assume an optically-thin,
unobscured spherical shell with total covering fraction $f_{\rm c}$.

It is clear from Table~3 that the inner warm absorber cannot contribute
much to the observed coronal line emission.  This is primarily due to the
fact that iron is too highly ionized, although collisional de-excitation is
also relevant in suppressing the coronal line emission.  On the other hand, the
outer warm absorber can produce significant coronal emission.  If we
hypothesize that all of the [FeX]$\lambda 6375$ and [FeXIV]$\lambda 5303$
emission originates from the outer warm absorber, we deduce that $f_{\rm
c}\approx 0.04$.  However, half of the optical emission from this region
may well be blocked by very optically-thick material (e.g. the molecular
torus).  This is suggested by the fact that, in some other Seyfert nuclei,
even infrared coronal lines are observed to be blueshifted with respect to
the low-ionization narrow lines implying that any redshifted coronal
emission must be heavily extinguished.  If this is the case, the true
obscuration-corrected coronal line flux maybe twice that
observed leading to a revised covering fraction of $f_{\rm c}\approx 0.08$.
This compares well with the estimate of the covering fraction of the outer
warm absorber, $f_{\rm owa}\sim 0.1$, based on the analysis of {\it ASCA}
data for a sample of Seyfert galaxies (Reynolds 1997).

According to our {\sc cloudy} calculations, the observed [FeXI]$\lambda
7892$ emission cannot be explained as originating from the same material as
the other coronal lines.  Within our hypothesis, three possibilities
present themselves.  First, the observed [FeXI]$\lambda 7892$ may have its
origins elsewhere.  Any separate coronal line emitting component would then
be heavily constrained by the fact that it could not over-produce
[FeX]$\lambda 6375$ and [FeXIV]$\lambda 5303$.  Secondly, the uncertainties
in the atomic physics may lead {\sc cloudy} to grossly underestimate the
[FeXI]$\lambda 7892$ flux and this line may, in fact, originate within the
same material as the other coronal lines.  Thirdly, the uncertainties in
the atomic physics may have led {\sc cloudy} to grossly overestimate the
[FeX]$\lambda 6375$ and [FeXIV]$\lambda 5303$ emission from the outer warm
absorber.  In this case, either the warm absorber covering fraction is
large ($f_{\rm c}\approxgt 0.1$) or the coronal lines are emitted from a
completely distinct (and as yet unidentified) region.  Further progress
in this area clearly requires better atomic parameters for these
transitions, such as those which will be provided by the {\sc iron} project
(Hummer et al. 1993).

\section{Further Discussion}

\subsection{Polarization}

Several authors have noted a high degree of polarization in the optical/IR
spectrum of MCG$-$6-30-15 (e.g. Thompson \& Martin 1988; Brindle et
al. 1990).  Polarization fractions of 4--6 per cent have been observed, with a
slight tendency towards increasing polarization with decreasing wavelength.
This is a significantly larger polarization than is typically found in
Seyfert nuclei.  The polarization is found to be aligned with the major
axis of the host galaxy.

There are two possible causes for this polarization (see Kartje 1995 for a
recent review of polarization mechanisms in AGN).  First, a large scale
alignment of dust grains along the line of sight to the central source can
produce polarization via dichroic extinction (i.e. one polarization is
preferentially scattered out of the line of sight by the dust grains).
Secondly, scattering of flux into the observers line of sight by dust or
free electrons will naturally produce polarization.  In MCG$-$6-30-15
we have independent evidence for both scattering and the presence of
line-of-sight dust.  Thus, both of these mechanisms may be relevant --
careful spectropolarimetry and detailed modeling will be needed to
disentangle these effects.

We note that IRAS~13349+2438, the other good dusty-warm absorber candidate,
also displays a very high degree of polarization (about 8 per cent).  Wills
et al. (1992) have found that the spectrum of the polarized light is
typical of an unreddened AGN, thereby suggesting that the polarization is
primarily due to scattering of the flux along a relatively dust-free path
into the observers line-of-sight.

\subsection{Global energetics}

Using the estimated unreddened spectrum, we can address the issue of the
energy distribution/flow in this source.   Here, we shall assume a
reddening of $E(B-V)=0.61$, leading to the unreddened spectrum sketched in
Fig.~4.   The (isotropic) luminosity in the 0.1--1000\,keV band is then
\begin{equation}
L({\rm X-ray}/\gamma{\rm -ray})\sim 2\times 10^{43}\ergps,
\end{equation}
the NIR/optical/UV luminosity is
\begin{equation}
L({\rm NIR/optical/UV})\sim 2\times 10^{43}\ergps,
\end{equation}
and the MIR/FIR luminosity is
\begin{equation}
L({\rm MIR/FIR})\sim 4\times 10^{43}\ergps.
\end{equation}
The NIR/optical/UV luminosity given here is really a lower-limit since we
have assumed the minimum possible value for the reddening.  These waveband
groups have been chosen on the basis of physical origin.  The
X-ray/$\gamma$-ray (i.e. $\sim 2\times 10^{16}\Hz$ and above) emission is
thought to be produced by non-thermal processes (e.g. Comptonization) in a
hot corona associated with the inner regions of an accretion disk.  The
NIR/optical/UV ($\sim 10^{14}-10^{16}\Hz$) emissions have plausible origins
as thermal emission from the optically-thick accretion disk material.  The
MIR/FIR ($\sim 10^{12}-10^{14}\Hz$) emission is likely to be thermal
emission from warm or hot dust associated with the dusty warm absorber and
putative molecular torus.  Here we address the implications of the relative
magnitudes of these luminosities for the energetics of the source.

First, we will discuss some theoretical expectations.  We will assume a
pure black hole model for the AGN emission, i.e. we will assume no
contribution to the observed `AGN' emission from a nuclear star-cluster or
starburst.  As mentioned in the Introduction, it is believed that the inner
accretion disk possesses a hot optically-thin corona responsible for the
non-thermal X-ray/$\gamma$-ray emission.  Coronal models coupled with
spectral constraints imply that a large fraction of the energy that is
(locally) dissipated in the accretion disk is transported into the
corona, possibly in a magnetic form, before being radiated.  The dominant
radiation process is thought to be inverse Compton scattering of soft
thermal (optical/UV) photons from the accretion disk.  The emission of the
optical/UV seed photons is probably driven by high-energy irradiation from
the corona, thereby completing a self-sustaining feedback.

Suppose that the corona covers the entire disk surface and that almost all
of the accretion energy is released within the corona leading to the
X-ray/$\gamma$-ray power-law emission.  Approximately half of the primary
high-energy photons will strike the accretion disk.  Approximately half of
the coronal\footnote{In this context, the term `coronal flux' is used to
mean that radiation which is associated with the X-ray emitting
disk-corona, and is unrelated to the `coronal line emission' discussed in
Section 5.} flux that strikes the disk will be thermalized and re-radiated
at optical/UV wavelengths, with the remaining half being `reflected'
(i.e. the photons undergo Compton backscattering or excite X-ray
fluorescence).  Thus, this scenario would predict
\begin{equation}
L({\rm NIR/optical/UV})=\Lambda L({\rm X-ray}/\gamma{\rm -ray}).
\end{equation}
where $\Lambda\sim 1/3$.  Observationally, we infer there to be
significantly more NIR/optical/UV emission than this, $\Lambda\approxgt 1$
(where approximate equality corresponds to the case where the reddening
takes its minimum allowed value, $E(B-V)=0.61$).  There are several
possible interpretations.  First, only a fraction of the (locally)
dissipated energy may be transported into the corona.  However, it is
difficult to reconcile this with the X-ray spectrum given current coronal
models (e.g., Haardt \& Maraschi 1991).  Secondly, there may be another
optical/UV source in addition to the accretion disk such as a powerful
nuclear starburst.  This is difficult to reconcile with the fact that the
optical continuum shown in Fig.~2a appears featureless and reddened to the
same degree as the BLR.  Lastly, and most likely, the corona may not cover
the whole disk.  It may be patchy or only exist in the innermost regions of
the disk.  The regions of the accretion disk without an active corona would
still produce optical/UV emission via thermal emission resulting from
viscous dissipation.

For the minimal reddening case, $E(B-V)=0.61$, the MIR/FIR luminosity is
comparable with luminosity in the whole of the rest of the spectrum.
Within the dust-reprocessing paradigm, this is a troublesome result to
understand unless the covering fraction of the dusty material is almost
unity or the primary emission is anisotropic (with more primary radiation
being emitted towards the dusty reprocessing material than towards us).  A
covering fraction of unity is implausible given our understanding of the
geometry of a Seyfert nucleus.  However, if we suppose that $E(B-V)>0.61$,
then the IR/optical/UV luminosity can greatly exceed the above value
thereby alleviating the problem of the MIR/FIR production.

\section{Conclusions}

We have presented a multiwaveband study of the Seyfert 1 galaxy
MCG$-$6-30-15, including previously unpublished optical data from the
AAT, and UV data from {\it IUE}.  Our compilation of data, spanning 6
decades of frequency, has allowed us to examine reprocessing mechanisms
and the geometry of this system.

The optical line and continuum emission both show the effects of dust
extinction.  The reddening inferred from Balmer line studies lies in the
range $E(B-V)=0.61-1.09$.  Given this reddening, we would expect the UV
emission from the source to be less than observed.  The fact that we do
detect a UV continuum and broad {\sc C\,iv}$\lambda 1549$ line can be
reconciled with a typical Seyfert spectrum if 1--5 per cent of the
intrinsic (i.e. dereddened) source spectrum is scattered around the matter
responsible for the extinction and into our line of sight.  UV
spectropolarimetry will be required to test this hypothesis.

The X-ray spectrum of this Seyfert nucleus clearly reveals a warm absorber
but little of the cold (neutral) absorption that would be expected to
accompany the dust responsible for the optical/UV reddening.  To reconcile
the X-ray absorption with the optical reddening we postulate that the dust
resides in the warm absorber.  Detailed X-ray studies have shown the warm
absorber to be comprised of at least two-zones -- an inner warm absorber at
distances characteristic of the BLR and an outer warm absorber at distances
characteristic of the putative molecular torus.  Dust cannot survive in the
inner warm absorber due to the intense radiation field -- thus, the dust
must reside in the outer absorber.  We have examined photoionization models
of the dusty (outer) warm absorber.  Upon comparing such models with {\it
ASCA} spectra we find evidence for non-standard dust mixtures, although the
current data to not allow us to pursue this issue in detail.

The optical spectrum of MCG$-$6-30-15 displays the high-excitation
forbidden lines of [FeX]$\lambda 6375$, [FeXI]$\lambda 7892$ and
[FeXIV]$\lambda 5303$, the so-called coronal lines.  As is often found in
other Seyfert nuclei, these lines have widths that are intermediate between
those of the broad Balmer lines and the lower-excitation forbidden lines.
Photoionization modeling with {\sc cloudy} suggests that the CLR may
be identified with the outer warm absorber.  

\section*{acknowledgments}

We thank Niel Brandt for several stimulating discussions over the course of
this work.  CSR acknowledges support from PPARC and the National Science
Foundation Grant AST9529175.  ACF thanks the Royal Society for support.
This work has made use of data obtained through the High Energy
Astrophysics Science Archive Research Center (HEASARC) Online Service,
provided by the NASA-Goddard Space Flight Center.

\end{document}